\documentclass{agujournal2019}
\usepackage{url} %this package should fix any errors with URLs in refs.
\usepackage{lineno}
\usepackage{booktabs}
\usepackage{multirow}
\usepackage{verbatim}
\usepackage{soul}
\newcommand{\responsetwo}[1]{#1}
\newcommand{\response}[1]{#1}

\newcommand{\model}{\textbf{DAGGER}}
\newcommand{\be}{\textbf{$\delta b_e$}}
\newcommand{\bn}{\textbf{$\delta b_n$}}
\newcommand{\sg}{SuperMAG}
\newcommand{\weimer}{\textbf{W2013}}

\draftfalse
\journalname{Space Weather}

\begin{document}

\title{Global geomagnetic perturbation forecasting using Deep Learning}

%% ------------------------------------------------------------------------ %%
%
%  AUTHORS AND AFFILIATIONS
%
%% ------------------------------------------------------------------------ %%

\authors{Vishal Upendran\affil{1,2}, Panagiotis Tigas\affil{1,3},  Banafsheh Ferdousi\affil{1,4} ,T\'eo Bloch\affil{1,5}, Mark C. M. Cheung\affil{1,6}, Siddha Ganju\affil{1,7}, Asti Bhatt\affil{1, 8}, Ryan M. McGranaghan\affil{1,9}, Yarin Gal\affil{1,3}}

\affiliation{1}{Frontier Development Lab, CA, USA}
\affiliation{2}{Inter University Centre for Astronomy and Astrophysics, Pune, India - 411007}
\affiliation{3}{OATML, University of Oxford,Oxford, UK}
\affiliation{4}{University of New Hampshire, Durham, NH, USA}
\affiliation{5}{University of Reading, Reading, UK}
\affiliation{6}{Lockheed Martin Advanced Technology Center, Palo Alto, CA, USA}
\affiliation{7}{NVIDIA Corporation, Santa Clara, CA, USA}
\affiliation{8}{SRI International, Menlo Park, CA, USA}
\affiliation{9}{ASTRA LLC, Louisville, CO, USA}
%(repeat as many times as is necessary)

\correspondingauthor{Vishal Upendran}{uvishal@iucaa.in}

%  List up to three key points (at least one is required)
%  Key Points summarize the main points and conclusions of the article
%  Each must be 100 characters or less with no special characters or punctuation and must be complete sentences

% Example:
\begin{keypoints}
\item	Global high-time cadence models for forecasting geomagnetic perturbations are necessary for this technologically-driven society.
\item	We develop a grid free model which forecasts these perturbations 30 minutes in the future at any spatial resolution at 1-minute cadence.
\item	The proposed model outperforms/has consistent performance against state of the practice local (global) high (low) time cadence models.
\end{keypoints}

%% ------------------------------------------------------------------------ %%
%
%  ABSTRACT and PLAIN LANGUAGE SUMMARY
%
% A good Abstract will begin with a short description of the problem
% being addressed, briefly describe the new data or analyses, then
% briefly states the main conclusion(s) and how they are supported and
% uncertainties.
%
%% ------------------------------------------------------------------------ %%

\begin{abstract}
Geomagnetically Induced Currents (GICs) arise from spatio-temporal changes to Earth’s magnetic field which arise from the interaction of the solar wind with Earth’s magnetosphere, and drive catastrophic destruction to our technologically dependent society. Hence, computational models to forecast GICs globally with large forecast horizon, high spatial resolution and temporal cadence are of increasing importance to perform prompt necessary mitigation. Since GIC data is proprietary, the time variability of horizontal component of the magnetic field perturbation (dB/dt) is used as a proxy for GICs. In this work, we develop a fast, global dB/dt forecasting model, which forecasts 30 minutes into the future using only solar wind measurements as input. The model summarizes 2 hours of solar wind measurement using a Gated Recurrent Unit, and generates forecasts of coefficients which are folded with a spherical harmonic basis to enable global forecasts. When deployed, our model produces results in under a second, and generates global forecasts for horizontal magnetic perturbation components at 1-minute cadence. We evaluate our model across models in literature for two specific storms of 5 August 2011 and 17 March 2015, while having a self-consistent benchmark model set. Our model outperforms, or has consistent performance with state-of-the-practice high time cadence local and low time cadence global models, while also outperforming/having comparable performance with the benchmark models. Such quick inferences at high temporal cadence and arbitrary spatial resolutions may ultimately enable accurate forewarning of dB/dt for any place on Earth, resulting in precautionary measures to be taken in an informed manner. 
\end{abstract}

% Key points
% 1. 
% 2. 

% 3. 

%% ------------------------------------------------------------------------ %%
%
%  TEXT
%
%% ------------------------------------------------------------------------ %%
\section*{Plain Language Summary}
Geomagnetically induced currents (GICs) result due to the interaction of the solar wind with Earth's magnetosphere, and are catastrophic to our technologically dependent society. Since GIC data is proprietary, the time variability of geomagnetic perturbation is used as a proxy, and forecasting these perturbation at high spatial resolution and time cadence is important. In this work we develop a deep learning based model to forecast these perturbation measurements at arbitrary spatial resolutions and at high time cadence, using only the solar wind measurements. Our model outperforms, or has consistent performance at worse with benchmark models, and hence can provide quick, accurate forecasts at high time cadence across the whole globe.

%--------------------------------
\section{Introduction}

Geomagnetic storms drive a spectrum of potentially catastrophic disruptions to our technologically dependent society~\cite{un_space}. A cohort study of insurance claims of electrical equipment provides evidence that space weather poses a continuous threat to electrical distribution grids via geomagnetic storms and geomagnetically induced currents~\cite<GICs;>{Schrijver:2014}. GICs also pose a threat to oil pipelines, railways, and telecommunication systems \cite{barlow1849, boteler-2001, pulkkinen-2001,eastwood2018quantifying}, potentially wiping out the backbone of economies and destroying the livelihoods of people worldwide. In the case of extreme but historically probable geomagnetic storms, the economic impact due to prolonged power outages can exceed billions of dollars per day ~\cite{Oughton:2017}. Hence, it is imperative to monitor and forecast space weather impacts like geomagnetic storms and GICs.

GICs are driven by the geoelectric field that depends on temporal changes in the horizontal component of ground magnetic field perturbation (dB/dt) and local Earth geology. Due to their proprietary nature, publicly available GIC data are limited.  However, the geomagnetic perturbations can be measured by Ground magnetometer stations, and  may been used as a good proxy to study GICs variations~\cite{gic_lanzerotti2001space,gic_kozyreva2018ground,gic_ngwira2018study}. The challenge, however, is two fold: (i). The ground magnetometers measurements are not performed uniformly across the Earth, and are spatially sparse (ii). Perturbation changes occur over timescales of minutes.

For predicting dB/dt, ground magnetic perturbations models at high spatial and temporal resolution are essential. Currently, first-principle models are used to forecast magnetic field perturbation as a part of the NOAA-SWPC (Space Weather Prediction Center) using Space Weather Modelling Framework~\cite<SWMF; see e.g.>{toth_2005_swmf,toth_2011_swmf,toth_2012_swmf}. The models generate forecasts of the global heliosphere, while they also provide forecasts of the magnetospheric parameters as a part of SWMF. However, these models are computationally expensive and require a long run time for high-resolution forecasts, which is necessary for highly localized magnetic field fluctuations. 

Data-Driven empirical models are more feasible for Space Weather forecasting due to their high speed and low computational cost \cite{camporeale-2019}. However, these empirical data-driven models \cite<e.g.>{weigel-2002,weimer2013empirical} did not perform well under Community-wide validation of geospace model ground magnetic field perturbation (dB/dt) predictions by \citeA{pulkkinen2013community}. The study was performed based on three first principles models and two empirical models as a function of upstream solar wind drivers using Heidke Skill Score (HSS) metrics over a number of ground magnetometer stations in mid and high latitudes. Further evaluation of the models by \citeA{welling-evaluations-2017} concluded all the models underpredict dB/dt during more active times and the need for model-data comparison and model improvements. 

Machine Learning (ML) and Deep Learning (DL) are rapidly growing areas that operate on large data. These have been used with great success in various studies -- right from forecasting the solar wind~\cite{upendran_2020_solarwind} to correlating auroral dynamics with Global navigation satellite system (GNSS) scintillations~\cite{lamb_2019_gnss}. \citeA{wintoft-dbdt-model} develop a neural network to forecast 30-minute maximum of $\vert$dB/dt$\vert$ at multiple stations over Europe with good success. More recently,~\citeA{Keese_2020_GICpaper} developed two models -- an artificial neural network model, and a Long Short Term Memory cell \cite<LSTM;>{hochreiter1997long} model -- to forecast the geomagnetic perturbations at the Ottawa station. While these studies forecast the perturbations at high temporal cadence (at $\approx$ 1-minute cadence), they are limited to forecast at specific spatial locations on the globe. 

In this work, we develop a \response{near} grid-free global geomagnetic perturbation forecasting model using deep learning to address the issues of near-real-time forecasts at high spatial and temporal cadence. \response{This is performed by coupling a DL model with spherical harmonic basis, rendering the model near grid-free.} The model takes the solar wind parameters, the Interplanetary Magnetic Field (IMF) measurements, and the solar radio flux measurements as input. It generates a forecast of perturbation measurements across the Earth with a lead time of 30 minutes. These forecasts may then be sampled over a grid at \response{nearly} any resolution. \response{Owing to the global nature of spherical harmonics, the perturbation forecasts may in-principle be sampled at any location on the globe.} The remainder of the paper is structured as follows:  in \S\ref{sec:data} we describe the data used in this work, along with the various pre-processing steps in \S\ref{sec:dataprep}. Then, we describe the main modeling scheme with the evaluation metrics in \S\ref{sec:metrics}, benchmark models in \S\ref{sec:bench}, and our proposed model {\model} ({\bf D}eep le{\bf A}rnin{\bf G G}eomagnetic p{\bf E}rtu{\bf R}bation ) in \S\ref{sec:FDL}. Finally, we present the results of our model in \S\ref{sec:results}, with detailed analysis on two selected storms in \S\ref{sec:stormperf}, and follow it up with a summary and broader impact in \S\ref{sec:summaryconclusion}.

%-------------------------
\section{Data}
\label{sec:data}

%-----------
\subsection{Perturbation measurement dataset}

In this study, we obtain the ground magnetic perturbations measurements from the {\sg}~\cite{gjerloev2012supermag} consortium. {\sg} is a global network of ground stations employed in measurement of geomagnetic perturbations. The available dataset comprises of measurements from around 300 magnetometer stations around the globe. These data are validated, transformed to a common co-ordinate system and processed with the same baseline remove methodology. From {\sg}, we obtain minute-resolution north-south perturbations in the geomagnetic field, {\be} and {\bn} from 2010 to 2019 at 1-minute cadence.

The {\sg} stations are primarily located in the Northern hemisphere. Thus, while the perturbations are densely sampled in the northern hemisphere, the sampling becomes sparser (and hence more susceptible to outliers) for lower latitudes and the southern hemisphere. \response{Particularly, the coverage of {\sg} stations is dense for MAGLAT $\ge40^\circ$. Hence, } to ensure a robust forecast and  as a first step in developing a grid-free model, we select stations only above a Magnetic Latitude (MAGLAT) of $40^\circ$. \response{Since we focus primarily on forecasting at MAGLAT  $\ge40^\circ$, our results are well constrained for the same regions. However, we emphasize that our solution formalism is generic enough to perform forecast anywhere on the globe -- the forecasts for regions with MAGLAT $\le40^\circ$ may however not expected to be as well constrained.}  This selection leaves us with a total of $175$ magnetometer stations (at max) to constrain our forecasts. 

%-----------
\subsection{Solar wind, IMF and solar proxy dataset}
We use the solar wind and IMF measurements \response{at 1-minute cadence from NASA/GSFC's OMNI dataset (through OMNIWeb)}. Particularly, we use: measurements of the three components of the interplanetary magnetic field (IMF) in GSM (Geocentric Solar Magnetospheric) coordinates ($\mathrm{B_x}$,$\mathrm{B_y}$,$\mathrm{B_z}$), solar wind speed ($\mathrm{V_{SW}}$), solar wind proton temperature ($\mathrm{T}$), the clock angle of the IMF($\theta_c$), and finally the the solar radio flux at 10.7 cm ($\mathrm{F}_{10.7}$) ~\cite{king_2005_omni,papitashvili2014omni}. 

From these basic measurements, we generate ``good'' features as input to our model following \citeA{weimer2013empirical}. We perform this feature generation to ensure accelerated convergence of our model, as these features are known to be important for reconstruction of the perturbation maps~\cite{weimer2010statistical,weimer2013empirical}.

The inputs to our model are: $\mathrm{B_x}$,$\mathrm{B_y}$,$\mathrm{B_z}$,$\mathrm{B_T}$, $\mathrm{V_{SW}}$, \rm{t} (dipole axis angle in radians), $\theta_c$, $T$, $\sqrt{\mathrm{F}_{10.7}}$, $\mathrm{B_T}\cos(\theta_c)$, $\mathrm{V_{SW}}\cos(\theta_c)$,  \rm{t}$\cos(\theta_c)$, $\sqrt{\mathrm{F}_{10.7}} \cos(\theta_c)$, $\mathrm{B_T}\sin(\theta_c)$, $\mathrm{V_{SW}}\sin(\theta_c)$, \rm{t} $\sin(\theta_c)$, $\sqrt{\mathrm{F}_{10.7}}\sin(\theta_c)$, $\mathrm{B_T} \cos(2\theta_c)$, $\mathrm{V_{SW}}\cos(2\theta_c)$, $\mathrm{B_T} \sin(2\theta_c)$, $\mathrm{V_{SW}}\sin(2\theta_c)$.

 %----------------------------------
\subsection{Data preprocessing}\label{sec:dataprep}
In general, the dataset for any machine learning work is split into three independent training, testing and validation sets. The training set is used to train the model, while the validation set is used to find the best model parameters which explain both the training and validation sets well. Finally, the model is evaluated on a testing set. Since we have a continuous time series of data which covers almost 75\% of the solar cycle, a naive division of different years into the three sets may result in bias due to prevalence of storms. Thus, in order to obtain a long enough time series to avoid edge effects, and mitigate bias from storm prevalence, we divide the whole time series into 100 buckets. Of these, we consider the two buckets with the 2011 and 2015 storms for benchmark. The remaining buckets are then split as 80\% training set, 10\% validation and 10\% testing set. \responsetwo{Also, note that following \citeA{weimer2013empirical}, we have included the F10.7 measurement, which is a widely used index of solar ultraviolet radiation levels and solar activity \cite{Verbanac_solarGeomagCorr,F107_Frederic}. While we expect F10.7 to provide some degree of information regarding the solar cycle, note that this index also shows localized variations~\cite{F107_tapping}. However, performing a detailed, quantitative analysis of the effect of the solar cycle on our model is beyond the scope of the current work. Hence, we may only expect some effect of the solar cycle to be captured by our model at this stage. }
 
The OMNI data at 1 minute cadence have missing values at multiple times, while the {\sg} measurements have missing data both at different times, and for different stations. Across the full dataset (train+test+val+storm), the OMNI solar wind measurements have the maximum missing data ($\approx25$\%). Similarly, for the two storm time series, the solar wind measurements again have the maximum number of missing data ($\approx18$\% for 2011, and $\approx24$\% for 2015 storm). The {\sg} data, on the other hand, have stations that go offline. This results in no target sample at the station location. During the storm times, the stations in consideration have a median missing fraction of $\approx5$\%. We report the median missing fraction for the missing {\sg} measurements as the missing stations do not contribute to our training scheme. 

To make the dataset uniform, we replace all missing values with 0 for both the OMNI and SuperMag data. To prevent any effect of missing measurements on our network, we replace the corresponding forecasts with 0 during training and validation time. This ensures that the ``error'' is zero for the particular sample, and that it does not contribute to training (and validation) of the network.
 
Before feeding the data (both OMNI and SuperMag) to our model, it is good practise to standardize the data by subtracting the mean and dividing by the standard deviation of the training set for each column. Due to memory constraints and the very large number of datapoints in the dataset, we generate the mean and standard deviation for 10,000 random points from the data. This ``Monte Carlo'' sample of points generates a mean and standard deviation which serves as a proxy for the training set mean and standard deviation. During inference time, these values are used to scale the validation and testing sets.
%-----------------------------------------------------
\section{Modelling and Methods}
\label{sec:modmeth}
%------------
\subsection{Metrics for model evaluation}
\label{sec:metrics}
We define multiple metrics to evaluate our model. For a target measurement of \rm{y} and forecast of $\hat{\mathrm{y}}$, the metrics are listed below:

\begin{enumerate}
    \item Root Mean Square Error (RMSE): \[ \mathrm{RMSE} := \sqrt{\frac{1}{N} \sum_{i}^N \left(\rm{y}-\hat{\mathrm{y}} \right)^2}, \] where the average is taken across all samples.
    \item Mean Absolute Error (MAE): \[ \mathrm{MAE} := \frac{1}{N} \sum_{i}^N \left(|\rm{y}-\hat{\mathrm{y}}|\right), \] where the average is taken across all samples.
\end{enumerate}

Apart from these two metrics, we also use the Pulkkinen-Welling metrics, which are based on binary event analysis for geomagnetic storms~\cite{Pulkkinen_metrics}. This analysis is performed only for the two storm series of 2011 and 2015, and not for the validation \& testing sets. For such an analysis, we define the horizontal perturbation component as \[ \delta b_{\mathrm{H}} = \sqrt{\be^2+\bn^2}, \]. The time derivative $d \delta b_i/dt$ is approximated as:

$$ \frac{d \delta b_{\mathrm{H},i}}{dt} \approx \sqrt{\left( \frac{\delta b_{e,i} - \delta b_{e,i-1}}{1 \mathrm{min} } \right)^2+\left(\frac{\delta b_{n,i} - \delta b_{n,i-1}}{1 \mathrm{min}}\right)^2}. $$
 
Following~\cite{Pulkkinen_metrics}, we divide our target and forecast {\be} and {\bn} for each station, into 20-minute non overlapping time windows. For each window, if $d \delta b_{H,i}/dt$ crosses a specified threshold, the segment is given a value 1 - else, it is given a value of 0. Thus, by comparing strings of 1s and 0s, we can then understand how good the model is at predicting events above or below a specific magnitude. \textbf{Hits} (H) are defined as number of correctly forecasted $1$s, while \textbf{Misses} (M) correspond to the number of measured $1$s marked  $0$ by the model. Similarly, \textbf{False alarms} (F) correspond to observed $0$s which are marked as $1$ by the model, while \textbf{True negatives} (N) are $0$s in the observation marked as $0$ by the model. Using this contingency table, we define four standard metrics following \citeA{welling_2018_metrics} to evaluate our model:
\begin{enumerate}
    \item Probability of Detection (POD): \[ POD = \frac{H}{H+M}. \]
    \item  Probability of False Detection (POFD): \[ POFD = \frac{F}{F+N}. \]
    \item Proportion Correct (PC): \[ PC =  \frac{H+N}{H+N+F+M}. \]
    \item Heidke Skill Score (HSS) is a measure of correctly predicted results after accounting for those which may be correct purely due to chance. The HSS is defined as: \[ HSS = \frac{2(HN-MF)}{(H+M)(M+N)+(H+F)(F+N)}. \]
\end{enumerate}
In this work, we select four different thresholds of 18, 42, 66, and 90 nT/min following \citeA{Pulkkinen_metrics}.

 %---------------------------------
\subsection{Benchmark models}
\label{sec:bench}
We use the 2011 and 2015 storm datasets, at 1 minute cadence as benchmark. Thus, the results presented here may be directly compared with other models evaluated on the same data~\cite<for example with the models proposed by>[]{Keese_2020_GICpaper}. However, we also have two self-consistent benchmark models operated on the same dataset.
 
The first, and the most simplest model is a persistence model. In our formulation, this model propagates the target {\sg} measurement at time T to T+LAG, where our LAG time is the forecasting horizon of our model. This propagation is performed for each station. Such a persistence model imposes a strong constraint on the utility of any proposed modelling scheme on ``how much'' new information is captured. For each target measurement, we also compute all the metrics for the persistence model. 
 
Our second benchmark model is the empirical fitting scheme of \citeA[henceforth called {\weimer}]{weimer2013empirical}. This is an empirical fitting scheme which decomposes the perturbation measurements into spherical harmonics, assuming the coefficients depend only on the solar wind parameters. This is a much stronger constraint over the persistence model for it actually generates a map between the solar wind and perturbation measurements. Note that the {\weimer} metrics are generated only for the two storm times, since we do not have the forecast for all times in our dataset.
%---------------------
\subsection{Proposed deep learning model: {\model}}
\label{sec:FDL}
The {\model} model is a deep learning model.\response{We use T hours of OMNI data at 1 minute cadence as input, and forecast the geomagnetic perturbations LAG minutes from the final input. The length of OMNI data and the LAG value are free parameters which are set through a hyperparameter search -- this is explained later in \S 3.4.} The model has three parts: a time series summarizer, a coefficient generator, and a spherical harmonic constructor. We first describe the spherical harmonic formulation, and then explain the full model.

% We use 120 minutes (2 hours) of OMNI data at 1 minute cadence as input, and forecast the geomagnetic perturbations 30 minutes from the final input. Thus the LAG value is $30$ minutes.
%----------
\subsubsection{Spherical harmonic formulation}\label{sec:sphharm}
Since we seek to develop a forecast model with continuous spatial coverage, we develop an almost ``grid free'' approach to forecast using Spherical Harmonics. Spherical harmonics assume a continuous and differentiable functional form of any field sought to be decomposed over a spherically symmetric manifold. Since we expect the perturbation fields to be largely smooth and devoid of localized peaks, we forecast the spherical harmonic coefficients, which can be easily transformed to the perturbations depending on the grid.

Any scalar field over the unit sphere can be expressed as \[ f(\theta,\phi) = \sum_{n=0}^{\infty} \sum_{m=-n}^{n} a_{nm} Y_{nm}(\theta,\phi), \] where $$Y_{nm}(\theta,\phi) :=  \sqrt{\frac{2n+1}{4\pi} \frac{(n-m)!}{(n+m)!}} e^{i m \theta} P^m_n(\cos(\phi)),$$ and $P^m_n(\cos(\phi))$ are the associated Legendre polynomials. 

These functions $Y_{nm}(\theta,\phi)$ are solutions to Laplace equation in a spherically symmetric coordinate system. If the sum is truncated at a maximum harmonic degree $N$, $f(\theta,\phi)$ is approximated as  
\begin{equation}\label{eqn:sphorig}
 \tilde{f}(\theta,\phi) = \sum_{n=0}^{N} \sum_{m=-n}^{n} a_{nm} Y_{nm}(\theta,\phi).
\end{equation}

Defining $i = n^2 + n + m$, we may rewrite Eq.~\ref{eqn:sphorig} as 
\begin{equation}\label{eqn:sph}
    f(\theta,\phi) \approx \tilde{f}(\theta,\phi) = \sum_{i=0}^{(N+1)^2-1} a_{i} Y_{i}(\theta,\phi).
\end{equation}

If the 2D fields over $\theta,\phi$ are unrolled as one-dimensional arrays, we have  \[ \tilde{f} = \mathcal{B} \vec{a}, \] where $\vec{a}=(a_i)$ is a vector of spherical harmonic coefficients, and  $\mathcal{B} = (\vec{b}_i)$ is the basis matrix wherein column vector $\vec{b}_i$ corresponds to the set of basis functions $Y_{nm}(\theta,\phi$).  \response{The maximum harmonic degree, or the number of modes $N$ is a free parameter which is fixed by hypterparameter tuning. This is explained in \S 3.4.}

We forecast both \be and \bn in this work. Hence, we generate coefficients for both the parameters with the same code.
%------------------------
\subsubsection{Model architecture}\label{sec:modarch}
We use a Gated Recurrent Unit (GRU) cell~\cite{cho2014learning} as a time series summarizer. A GRU cell is a variant of the Recurrent Neural Network~\cite{rumelhart_1985_rnn}. The GRU cell has an internal memory in the form of a ``hidden state'' which is updated as inputs are given to it. This update happens through a sequence of non linear projection and shift operations~\cite<see>[for details]{cho2014learning}. Thus, the input time series is used to update the hidden state, encoding the information content of the input time series. 

\response{We feed in the T hours of the solar wind measurements to the cell, which are summarized into a ``hidden state'' of the cell. Note again that this length of the time series is fixed through the hyperparameter search described in \S 3.4.} The hidden state vector has a size of 8 units. This state vector acts as a proxy for all the solar wind information needed for our forecast.

The hidden state is then fed into a fully-connected layer, which transforms the hidden state to a vector of coefficients. The number of coefficients is determined by the largest mode we seek to forecast from the code.

%-----------------
\begin{figure}[ht!]
    \centering
    \includegraphics[width=\linewidth]{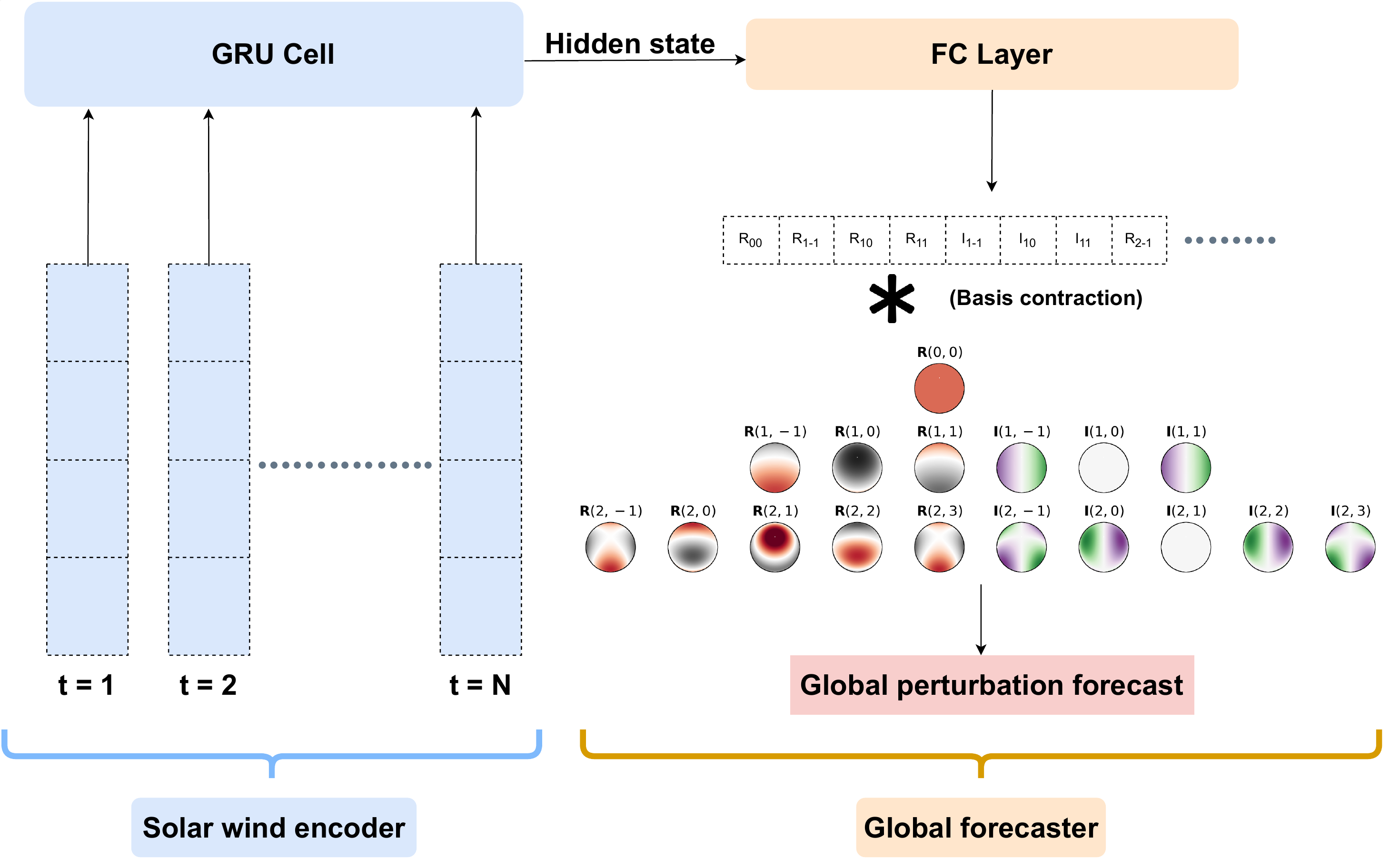}
    \caption{Architecture of {\model}. The model has three principle components -- a time series summarizer, a coefficient generator and a spherical harmonic constructor. The time series summarizer (GRU cell) takes in the solar wind time series, and generates a summary hidden state. This is fed to a fully connected layer (FC Layer), which generates a vector of coefficients. These coefficients are contracted with the spherical harmonic basis to generate global forecast of perturbations.} 
    \label{fig:model_arch}
\end{figure}
%------------------
\begin{table}[ht!]
	\centering
	\caption{Model architecture: A summary}
	\label{tab:model}
	\begin{tabular}{l l}
		\toprule
		\textbf{Layer name} & \textbf{Size}\\
		\midrule
		GRU & 8 units\\
		FC: MLP Layer 1 & 16\\
		FC: MLP Layer 2 & 440*2 (real and imaginary parts)\\
		Spherical harmonic layer (NOT trainable) & -- \\
		\bottomrule
	\end{tabular}
\end{table}
%----------------------

Finally, the output from the fully connected layer is then contracted with the spherical harmonic basis, giving out the forecast of perturbation measurements at any required spatial location. \response{This basis, which enforces our GRU hidden state to be the spherical harmonic coefficients, is called the Spherical harmonic basis layer. Since the basis functions are computed using their analytical form (and not learned through data), the basis layer is not trainable.} The model architecture is summarized in Fig.~\ref{fig:model_arch}, while the layer sizes are provided in Table.~\ref{tab:model}.

The MLT of various {\sg} stations change with time. Hence, during training and inference time, the $\mathcal{B}$ are evaluated during every forward pass for the (MAGLAT,MLT) of the stations where the measurements are made. Hence, the spherical harmonic coefficients are constructed during each forward pass. Also note that the spherical harmonic formulation presented in Sec.~\ref{sec:sphharm} has the azimuth origin at the North pole. Hence, we transform the MAGLAT into Magnetic co-latitude. 
%-----------------------
\subsection{Hyperparameters}\label{sec:hyperparam}

Deep learning models generally have trainable parameters (which we shall henceforth call weights), and free parameters which must be set manually (called hyperparameters). We monitor the validation set performance for different combinations of the hyperparameters, and use a Bayesian grid search to select the hyperparameters which give the best validation set performance as the final model. \response{A Bayesian grid search is a more informed search over a random search, which updates the next to-be-tested hyperparameter combination conditioned on the previous samples and validation set performance.} We performed the hyperparameter search using Weights and Biases~\cite{wandb}. The hyperparameters values are given in Table.~\ref{tab:hyperparam}. \response{The hyperparameter grid or bounds of the distributions are provided in the Supplementary section.}

%----------------------
\begin{table}[ht!]
	\centering
	\caption{Hyperparameters set through Grid search.}
	\label{tab:hyperparam}
	\begin{tabular}{l l}
		\toprule
		\textbf{Hyperparameter} & \textbf{Value}\\
		\midrule
		OMNI time series length & 120 min \\
		Maximum number of modes & 20 \\
		Learning rate & $5\times10^{-3}$ \\
		L2 regularization coefficient & $5\times10^{-5}$ \\
		Dropout probability & 0.7 \\
		Batch size & 8500 \\
		Optimizer & Adam, with default Pytorch parameters\\
		\bottomrule
	\end{tabular}
\end{table}
%----------------------

With the model hyperparameters and architecture fixed, we train the model. We use the Mean Absolute Error (MAE) as the loss function to be optimized. The L2 regularization is a penalization term preventing the coefficients from growing too large. This penalty term serves the two-fold benefit of preventing overfitting, and reducing sparsity amongst the coefficients. Since we would want as many harmonics to be captured as possible to better resolve local disturbances, we would want the ``power'' to be spread across as many modes as possible. Furthermore, we use dropouts~\cite{srivastava2014dropout} to randomly switch off neurons during the training time to enhance independent pathways within the model. This again serves to prevent overfitting in the model. 

As already mentioned in Sec.~\ref{sec:data}, we train the model on $\approx$ 10 years of data, and report the results below. We performed the training on an NVIDIA A100 GPU with 40 GB memory, with the model taking $\approx$ 40 hours for convergence. 
%-----------------------------------------------------
\section{Results}\label{sec:results}
\subsection{Testing set}\label{sec:testingset}
%-----------------
We report the results, and performance of our model below. This is done in two ways: first, we report the statistics on the test set, and next we report the performance of our model for the 2 storm times described in \S \ref{sec:data}. For both the datasets, we benchmark our model agains the persistence model, while for the storm times, we also benchmark against {\weimer}. Furthermore, we also report the event-based metrics for the two storm times, to enable comparison across other models and papers.

%------------------
\begin{figure}[ht!]
\centering
\includegraphics[width=\linewidth]{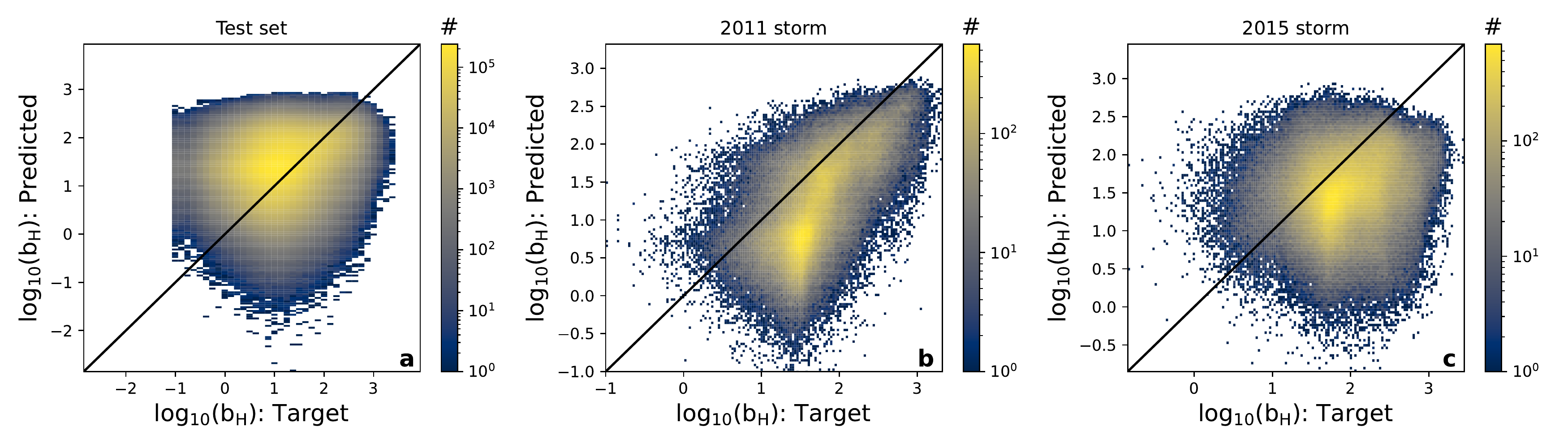}
    \caption{The joint histogram of predictions v/s target SuperMag measurements for all points in the test set (panel \textbf{a}), 2011 storm set (panel \textbf{b}) and 2015 storm set (panel \textbf{c}). The colours depict the number of points in each bin of the joint histogram.}
    \label{fig:scatter}
\end{figure}
%------------------

In Fig.~\ref{fig:scatter}, we report the joint distributions of the forecast and the target $\delta\mathrm{b}_{\mathrm{H}}$ for the test set (panel \textbf{a}), 2011 storm set (panel \textbf{b}) and 2015 storm set (panel \textbf{c}). Since there are a large number of points, the number of points in each bin is shown using the colorbar. Note that the number points (and hence the colour) scale logarithamically. The black line shows a slope=1 line. For a perfect forecast, all points should lie on this line -- however, this is seldom the case. From Fig.~\ref{fig:scatter}, we see that by and large the model predictions and targets are aligned to the slope=1 line. Furthermore, the 2011 storm is better forecasted than the 2015 storm. \response{However, note that we can also see that {\model} also has a tendency to ``under-forecast'', since there exist more points below the line slope=1 than above.}

On the held-out testing set, we obtain an RMSE (MAE) of $35.28\mathrm{nT}$ ($20.41\mathrm{nT}$) and $63.74\mathrm{nT}$ ($39.36\mathrm{nT}$) for \be and \bn respectively. Hence, we see a clear effect of outlier datapoints in the computation of these metrics, resulting in higher value of RMSE over MAE. For the persistence model, we obtain an RMSE (MAE) of 26.46 (10.39) nT and 35.88 (13.63) nT for \bn and \be respectively. Thus, while our model shows low errors, it does not quite beat the persistence model in these metrics. Hence, significant ``autocorrelation'' of the perturbations seems to exist within a forecast horizon of 30-minutes, which results in the low RMSE and MSE of persistence model. 

However, RMSE and MAE do not quite give us any information regarding the temporal structure of the forecasts with respect to our measurements. Hence, we next validate our model performance across the two storm datasets. For these two storms, we have the nowcast from {\weimer} and the persistence model to benchmark our performance.
%--------------------
\subsection{Storm time performance}\label{sec:stormperf}

We now report the RMSE and MAE of our forecasts and the benchmarks, for the two storm datasets in Table.~\ref{tab:msemae}. Note that the metrics are calculated across all times and all stations. From Table.~\ref{tab:msemae}, we again clearly see the feature of larger RMSE over MAE due to outlier cases in the dataset. The persistence model shows metrics only marginally better than {\model} forecasts for the 2011 storm -- infact, the RMSE in {\be} is lower for {\model}. However, this is not the case for the 2015 storm. Hence, a 30-minute time window still contains significant auto-correlation in the {\sg} measurements, as we have also seen from the testing set results. 

{\model} clearly outperforms {\weimer} in both the metrics for both the components of the horizontal magnetic field perturbation. Since the primary input features to our model are the same as those used by {\weimer}, these results tell us that deep learning is able to capture a much more non-linear association between the solar wind/IMF/solar flux and geomagnetic perturbation measurements.

\begin{figure}[ht!]
    \centering
    \includegraphics[width=\linewidth]{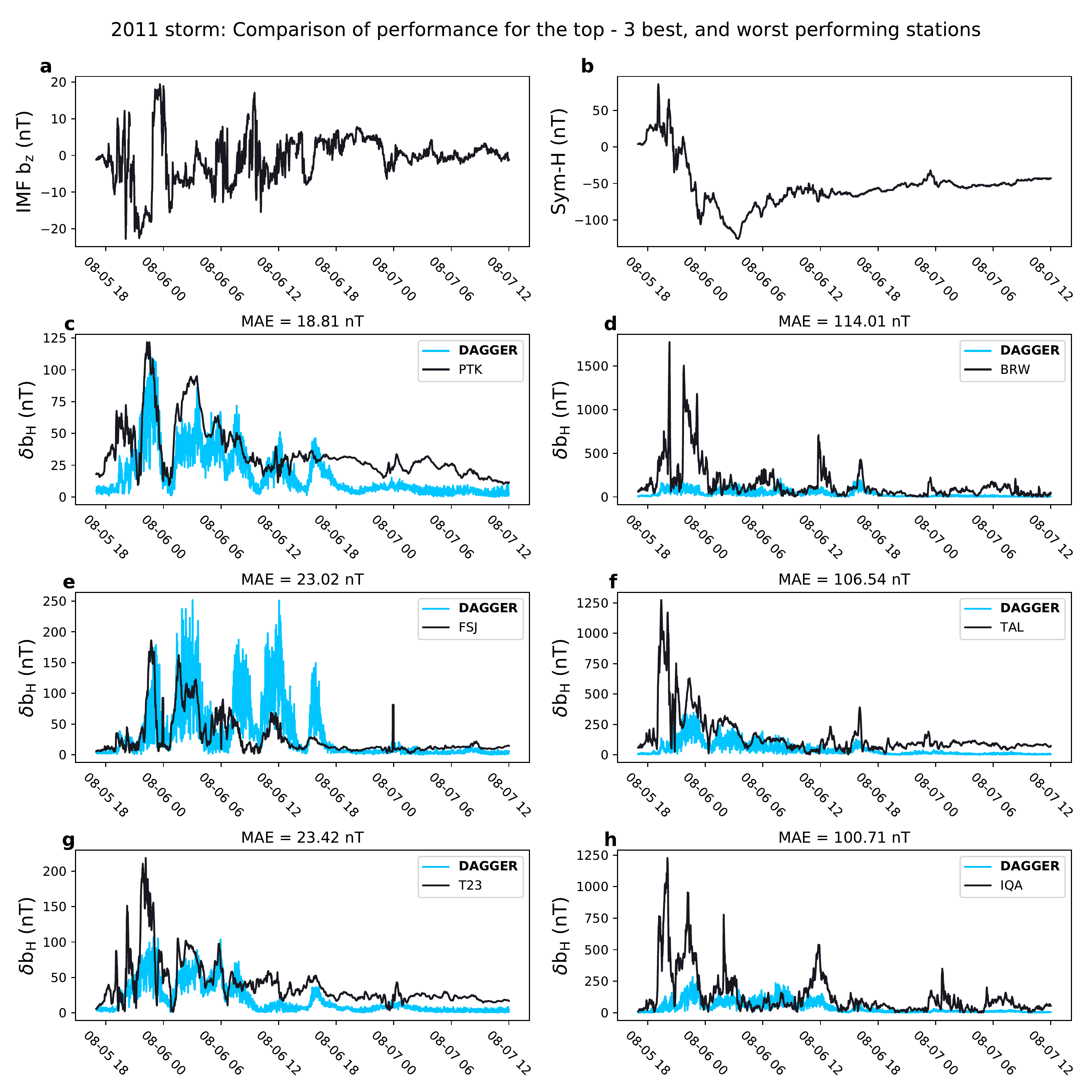}
    \caption{The IMF Bz (panel \textbf{a}), Sym-H (panel \textbf{b}) and top 3 best (panels \textbf{c}, \textbf{e}, \textbf{g}) and worst (panels \textbf{d}, \textbf{f}, \textbf{h}) performing stations for the 2011 storm. The blue colour indicates forecast from {\model}, while the black colour indicates measurements at different stations (in the legend of each figure), with the MAE reported on top.}
    \label{fig:top3_2011}
\end{figure}

\begin{figure}[ht!]
    \centering
    \includegraphics[width=\linewidth]{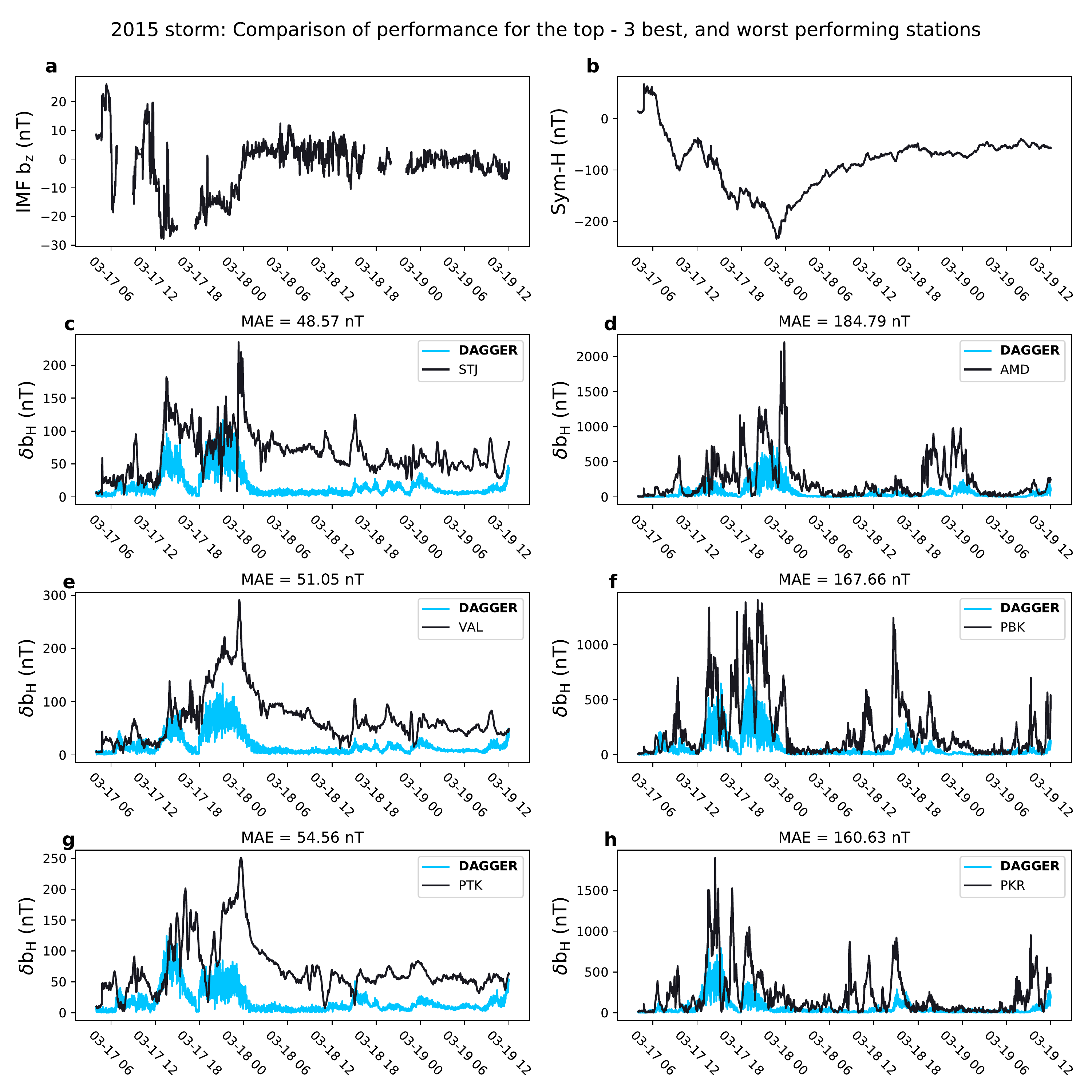}
    \caption{Same as Fig.~\ref{fig:top3_2011}, but for the 2015 storm}
    \label{fig:top3_2015}
\end{figure}

%--------------------
\begin{table}[ht!]
\centering
\caption{RMSE and MAE comparison between {\model}, {\weimer} and Persistence models. Both the metrics are in units if nT.}
\label{tab:msemae}
\begin{tabular}{cc|cc|cc|cc}
    \cline{1-8}
    \multirow{2}{*}{Storm} & \multirow{2}{*}{Metric} &   \multicolumn{2}{c|}{{\model}} & \multicolumn{2}{|c}{{\weimer}}
    & \multicolumn{2}{|c}{Persistence }\\
    \cline{3-8}
    & & \be & \bn & \be & \bn & \be & \bn \\
    \cline{1-8}
    \multirow{2}{*}{2011} & MAE & 34.99 & 53.20 & 67.41 & 76.74 &30.87 &43.52 \\
    & RMSE  & 72.86 & 100.46 & 127.54 & 140.93& 73.53 & 97.41 \\
    \cline{1-8}
    \multirow{2}{*}{2015} & MAE & 61.44 &  104.7 & 104.69 & 121.48 & 47.17 &67.4 \\
    & RMSE  & 102.45 &  \response{175.37} & 179.97 & 195.52 & 87.78 & 128.90\\
\cline{1-8}
\end{tabular}
\end{table}
%------------------
To investigate ``how good'' RMSE and MAE are as metrics to quantify performance, we present the forecasts from our model and compare them with the measurements at different stations. Since there are $\le175$ stations in our dataset, we present results for the ``best'' and ``worst'' forecasted stations. To this end, we select forecasts for 3 stations which show the smallest, and largest MAE. Every other forecast would lie somewhere between the best and worst case scenarios. These forecasts for the two storms, along with IMF Bz and Sym-H indices are shown in Fig.~\ref{fig:top3_2011} and \ref{fig:top3_2015} for the two storms. 
%--------------------------------
%-------------------------------
\begin{figure}[ht!]
    \centering
    \includegraphics[width=\linewidth]{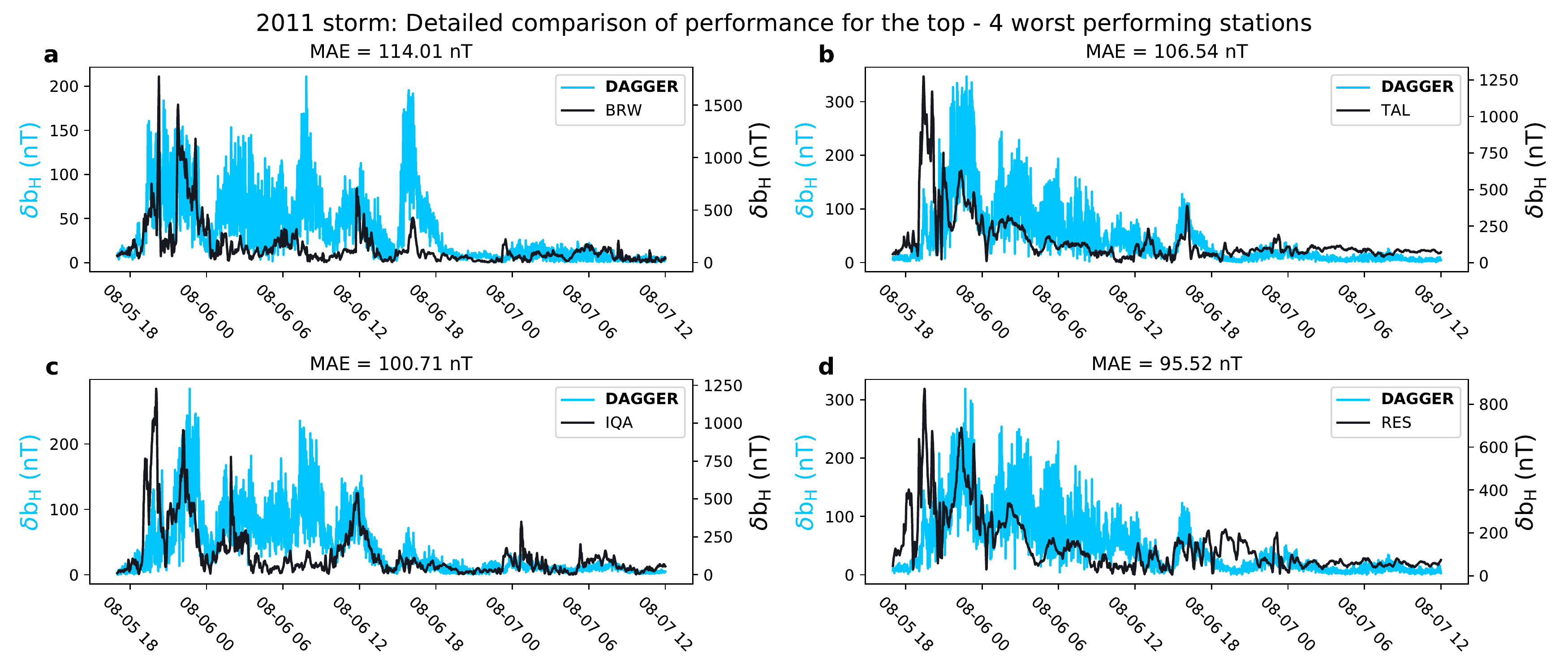}
    \caption{The measurements at different stations (black) and forecast (blue) of {\model} for the 2011 storm, with different Y-axis scales to bring out the detailed features from Fig.~\ref{fig:top3_2011}.}
    \label{fig:top3_2011_detailed}
\end{figure}
%-------------------------------

From Fig.~\ref{fig:top3_2011}, we see that the dichotomy between the best and worst performing forecast is quite stark. First, we clearly see that the forecasts deemed ``best'' (left column) correspond to stations where the measurements are $\le250$ nT. Second, {\model} is able to clearly pick out the different peaks and troughs of the forecast -- especially for the stations with lowest MAE (panel. \textbf{c}). Third, the perturbation forecast and measurement values are of similar magnitudes, and in many cases match well for the stations which show the lowest errors. On the other had, prima-facie it looks like {\model} is unable to forecast anything at all for the stations with large MAE. Clearly, the largest perturbation measurements from these stations are $\approx6\times$ the largest perturbations for the stations showing low MAE. Since some salient variations seem to be captured by {\model} (see panelsf and h), we define different Y-axes for the forecast and measurement, to probe how goof (or bad) {\model} forecasts for these stations in Fig.~\ref{fig:top3_2011_detailed}.

From Fig.~\ref{fig:top3_2011_detailed}, two inferences may be made. First, {\model} is able to forecast the variation of perturbation over time even for stations which have a large associated MAE. And second, {\model} under-predicts the large perturbation values, which hence gives rise to a large MAE. Thus, a purely deep learning framework is able to assimilate the solar wind measurements and generate salient associations with the magnetic field perturbations. The exact scale of perturbations is however missed for the stations with large associated MAE. 
%-----------------------
%-----------------------
\begin{figure}[ht!]
    \centering
    \includegraphics[width=\linewidth]{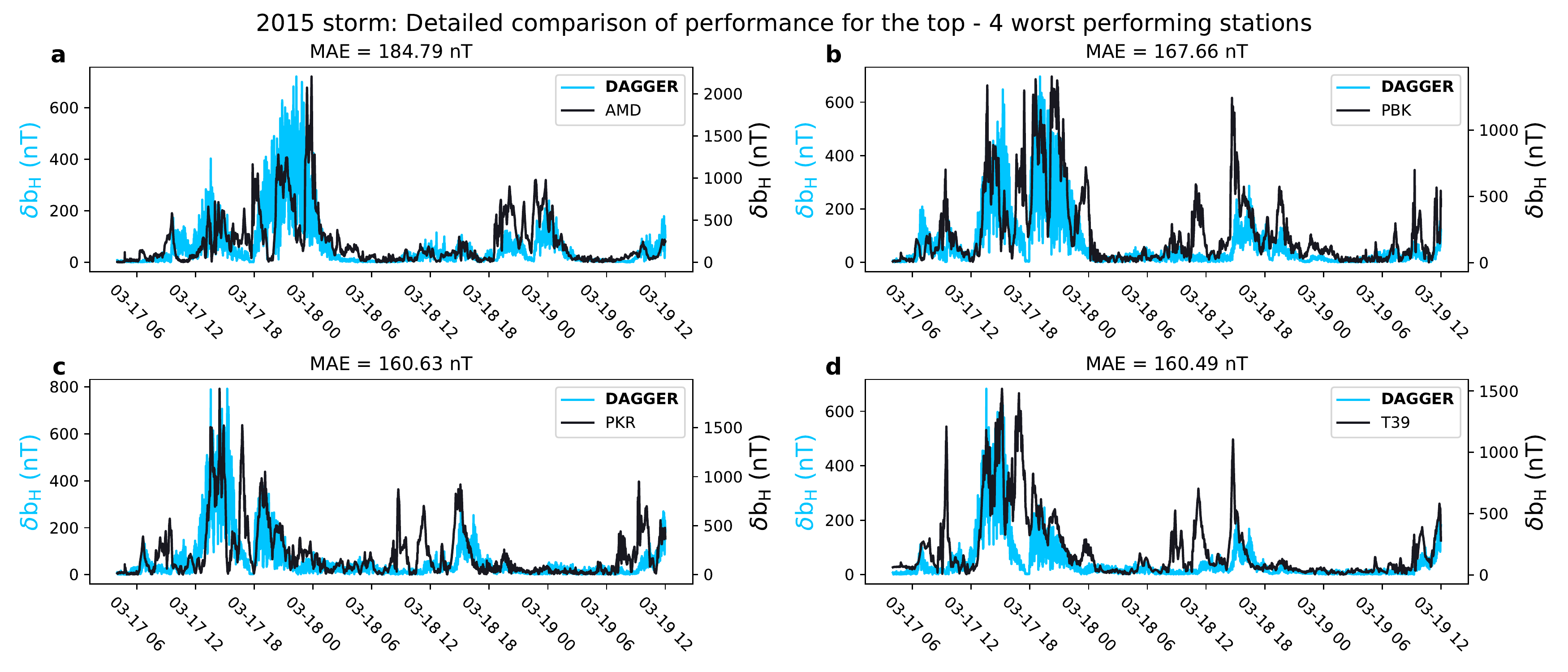}
    \caption{Same as Fig.~\ref{fig:top3_2011_detailed}, but for the 2015 storm.}
    \label{fig:top3_2015_detailed}
\end{figure}
%-------------------------------

These results are also clearly seen in Fig.~\ref{fig:top3_2015} for the 2015 storm dataset. The stations having low MAE typically have a max perturbation of $\approx300$ nT, while the stations with largest MAE are $\approx6\times$ larger. One interesting result to be noticed for the stations with lowest MAE for this storm is the mismatch between forecast and measurement is larger for the 2015 storm than in the 2011 storm case (compare panels \textbf{c}, \textbf{e} and \textbf{g} between Fig.~\ref{fig:top3_2011} and Fig.~\ref{fig:top3_2015}). This is also consistent with larger spread in the joint histograms in Fig.~\ref{fig:scatter}.\textbf{c}. To see if this is also observed for the stations with large MAE, we check the forecast and measurements on different Y-axis scales in Fig.~\ref{fig:top3_2015_detailed}.

From Fig.~\ref{fig:top3_2015_detailed}, we once again see that {\model} is able to capture salient variation of the perturbation measurements, but fails to reproduce the exact values. However, both the lowest and largest MAE for the 2015 storm are larger than those for the 2011 storm. From Fig.~\ref{fig:top3_2015}.\textbf{a} (and also from \S\ref{sec:dataprep}), we see that the 2015 storm measurements have a lot of data gaps. This is not seen for the 2011 storm (see Fig.~\ref{fig:top3_2011}.\textbf{a}). Hence, we speculate that the larger MAE for the 2015 storm arises from a lack of data (which may also depend on the imputation scheme), resulting in spurious forecast when the solar wind data is missing.

\begin{figure}[ht!]
    \centering
    \includegraphics[width=0.3\linewidth]{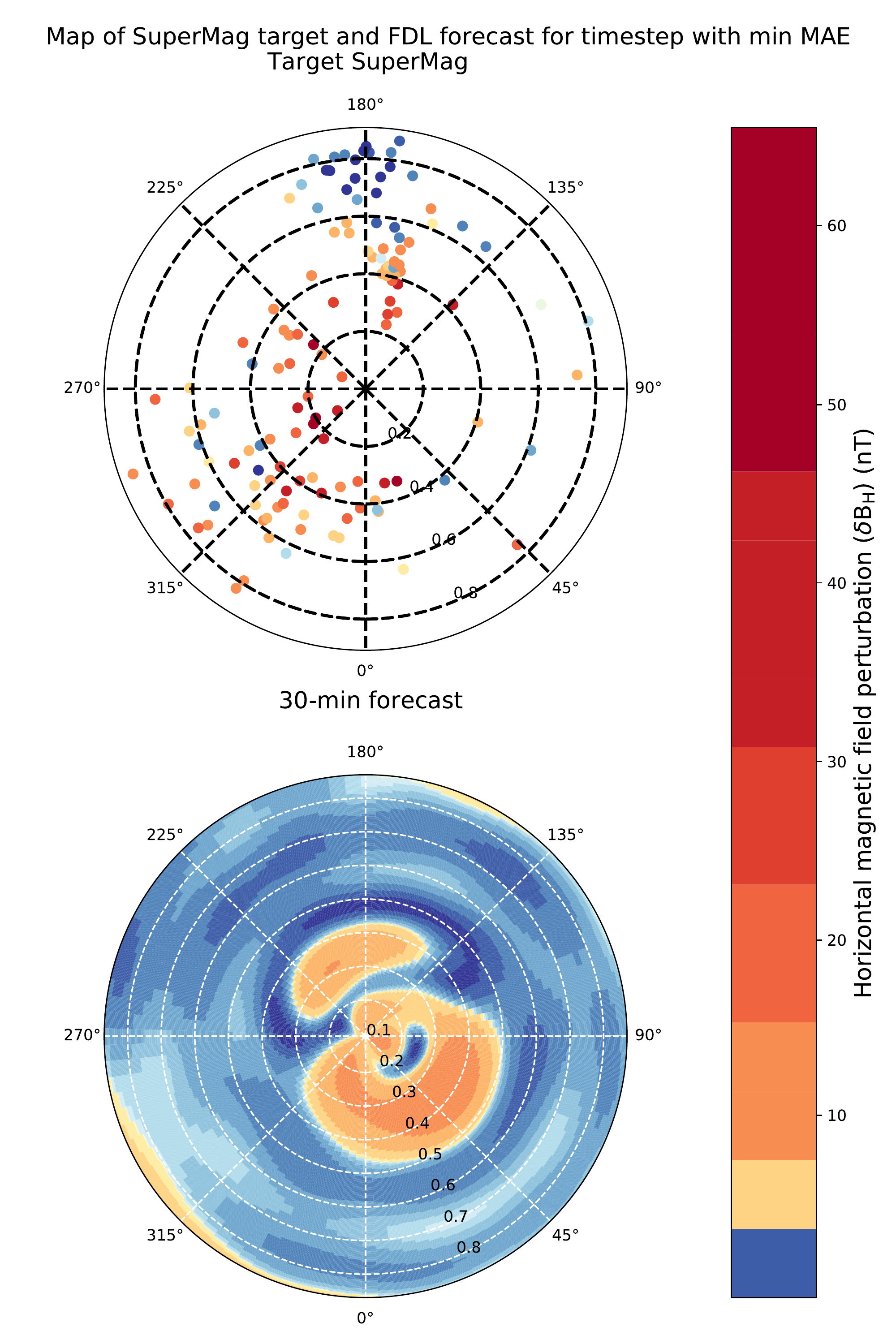}
    \includegraphics[width=0.3\linewidth]{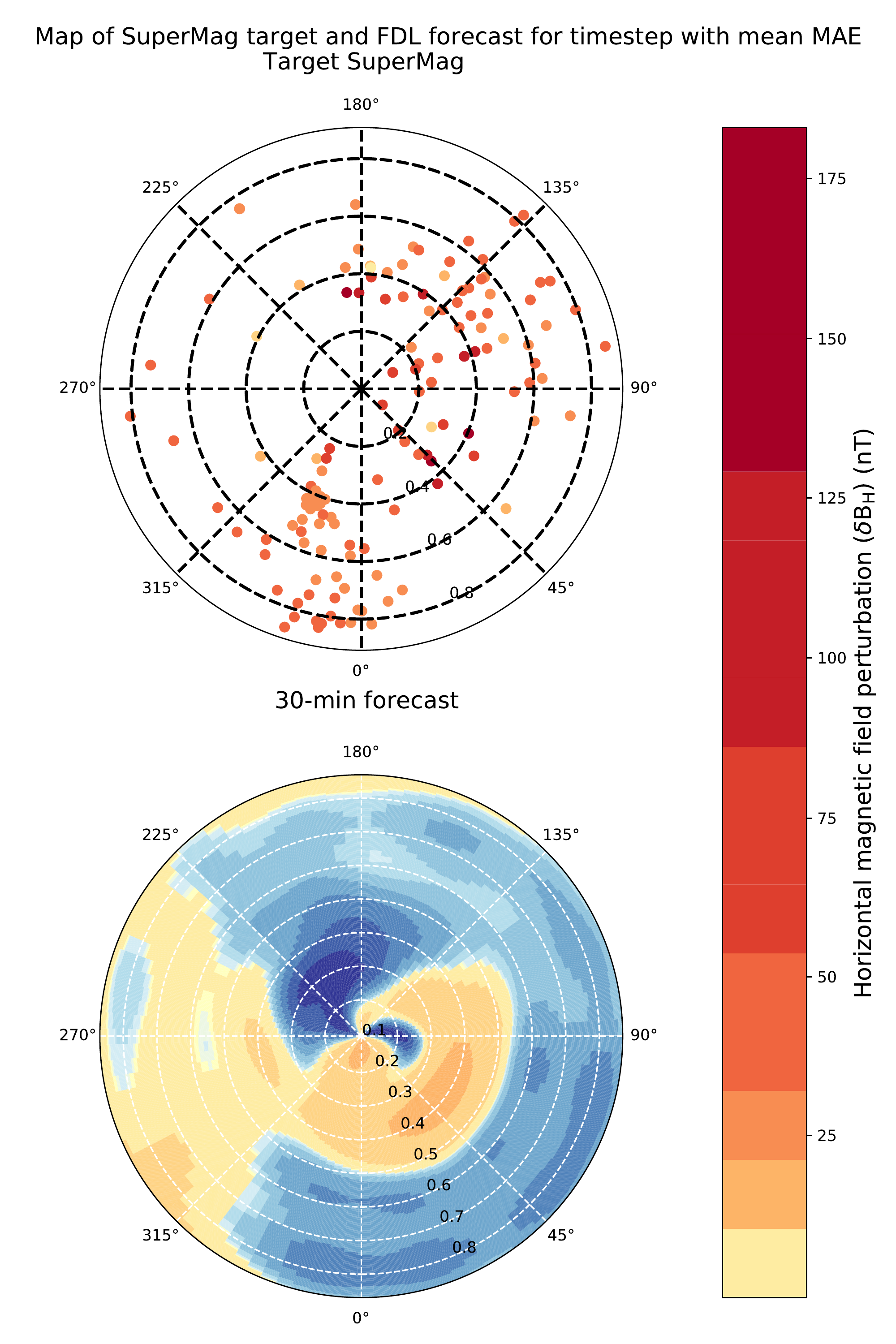}
    \includegraphics[width=0.3\linewidth]{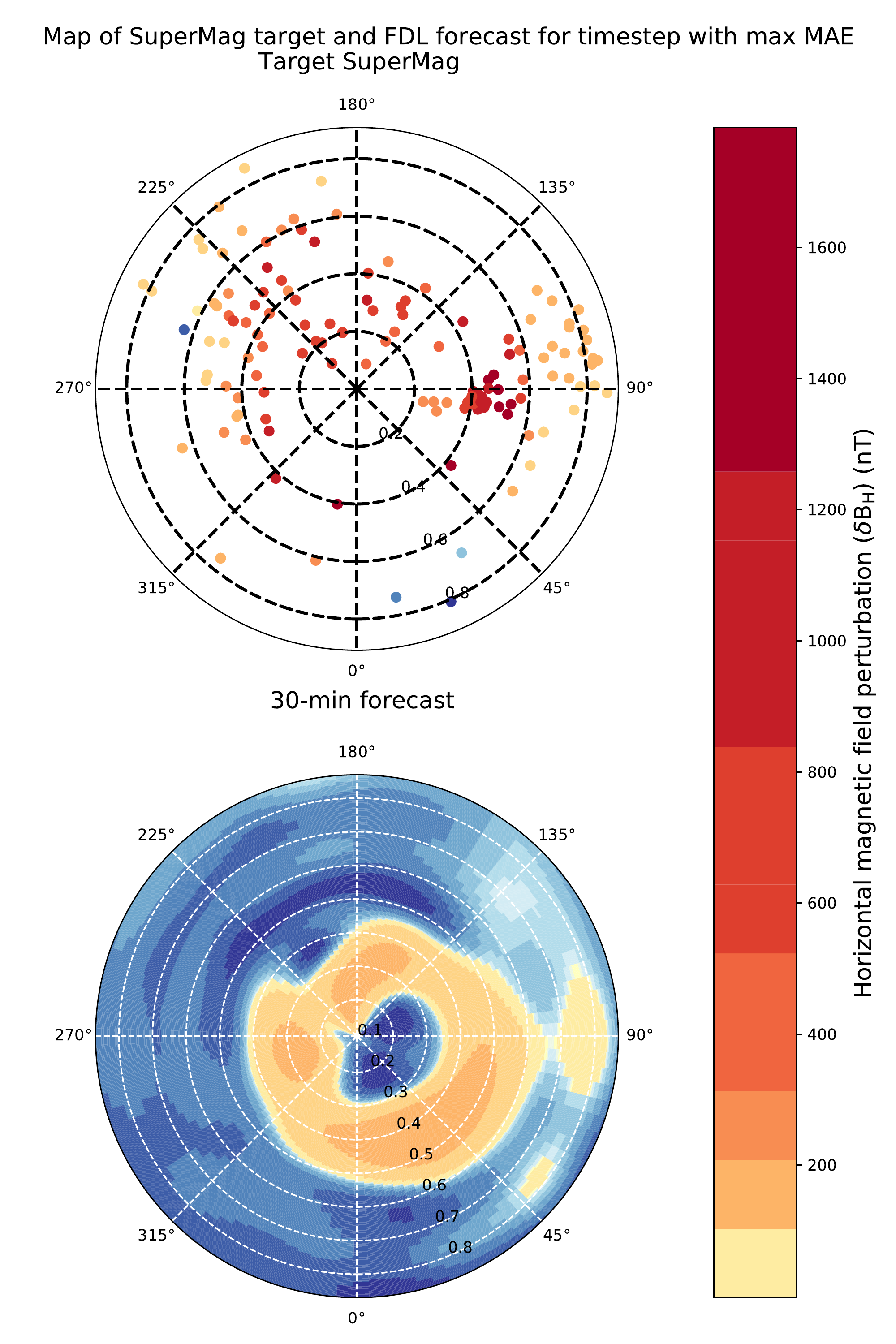}
    \caption{Maps of the measurement (top row) and forecast (bottom row) for the 2011 storm at times with minimum (left), mean (center) and maximum (right) MAE. }
    \label{fig:global_2011}
\end{figure}

\begin{figure}[ht!]
    \centering
    \includegraphics[width=0.3\linewidth]{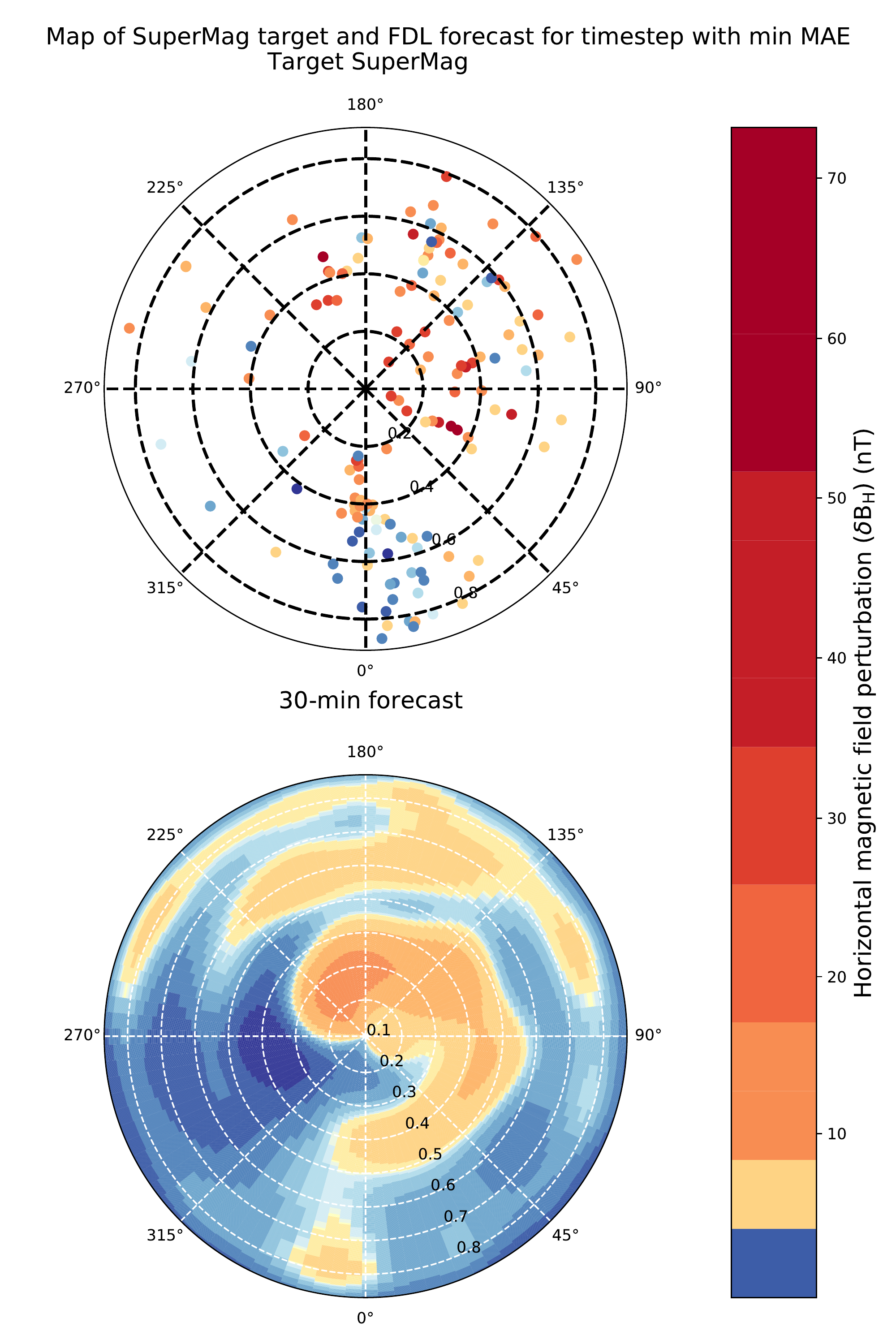}
    \includegraphics[width=0.3\linewidth]{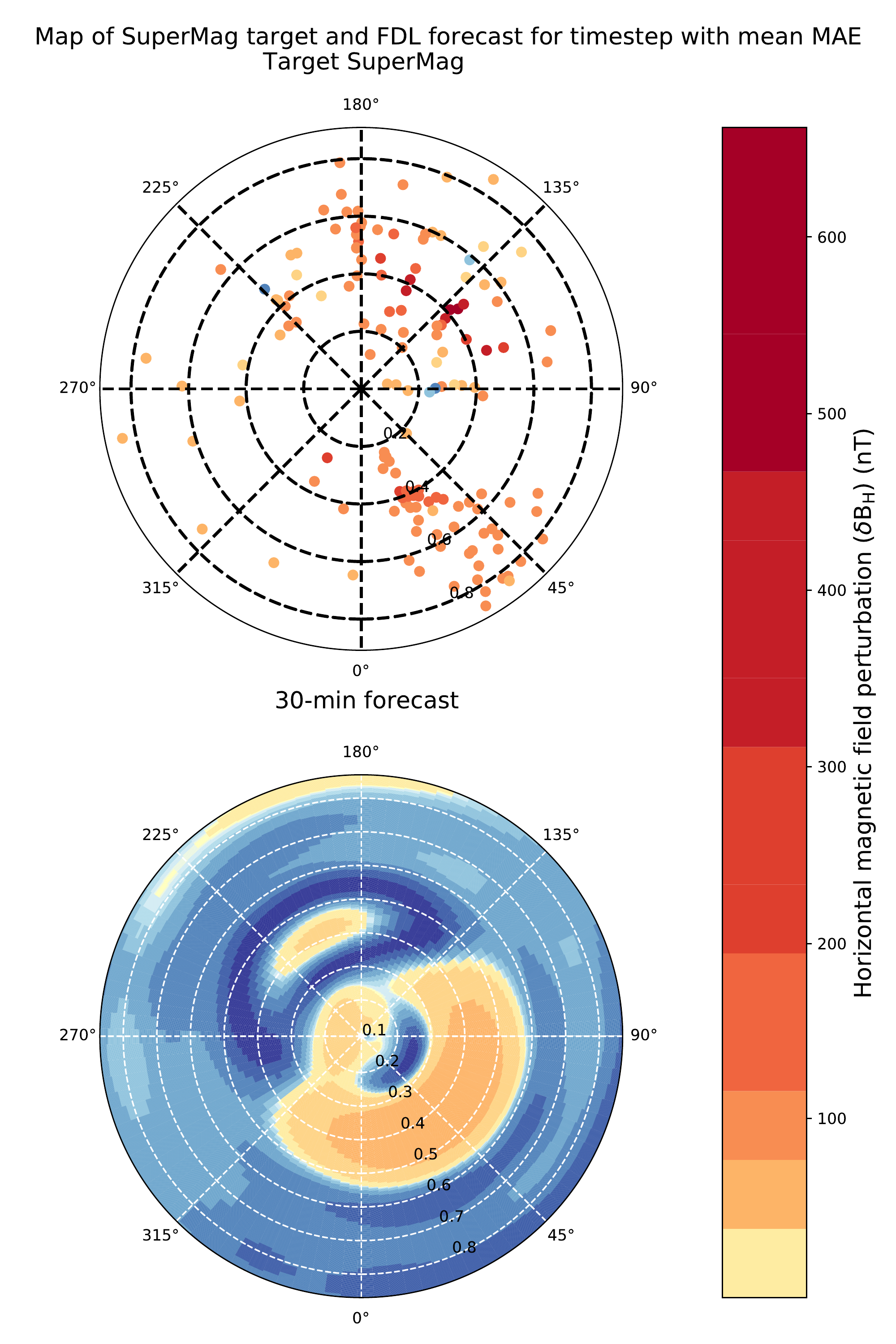}
    \includegraphics[width=0.3\linewidth]{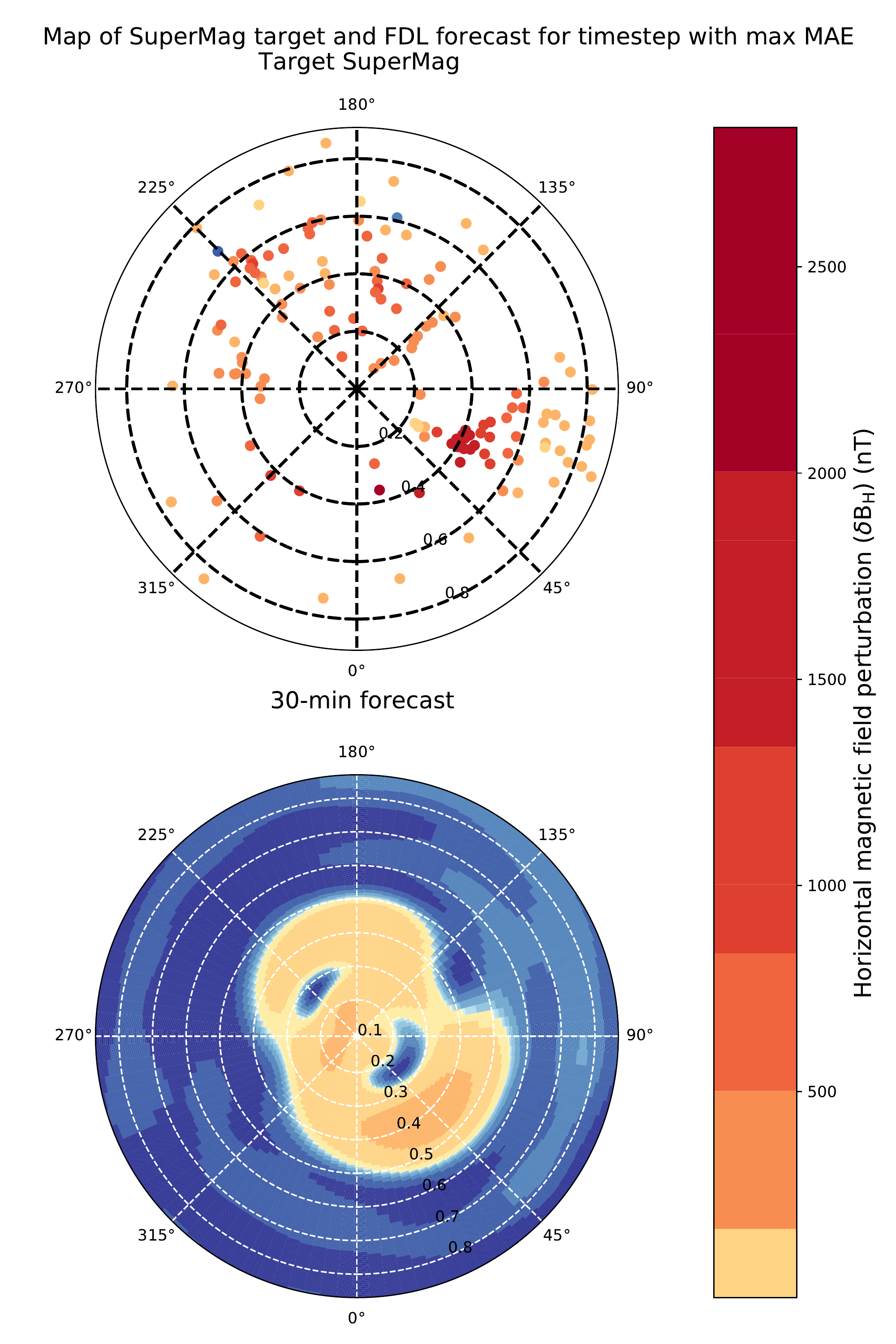}
    \caption{Same as Fig.~\ref{fig:global_2011}, but for the 2015 storm. }
    \label{fig:global_2015}
\end{figure}

In Fig.~\ref{fig:global_2011} and Fig.~\ref{fig:global_2015}, we show the maps for $\delta\mathrm{b}_{\mathrm{H}}$ (forecast in the bottom row, perturbations in the top) in MLT-MCOLAT grid, with the center being the North pole. This is done for 3 cases -- this time, for the timestep with minimum, mean and maximum MAE across all stations. From these plots, we clearly see that our model provides a dynamic map of the perturbations, at a cadence of 1 minute. Furthermore, it also shows how under-prediction gives rise to the larger MAEs. Thus, such perturbation maps for {\bn}, {\be} and $\delta b_{\mathrm{H}}$, changing dynamically over time scale of $\approx1$ minute are made available across the two storm times as video files in the online version of the paper.
%-------------------------------
%----------------------
%-----------------------

Finally, we clearly see that MAE (or MSE) are good detectors of magnitude-match of the forecast with the measurements, but cannot pick out if the variations are captured with specific thresholds. Thus, we also present the event based metrics as a measure of {\model} performance in Table.~\ref{tab:evenbasedmet}.

In Table~\ref{tab:evenbasedmet}, the metrics are computed for each station, and we report the mean and standard errors across all the stations. The standard error is defined as $\sigma/\sqrt{N}$, where $\sigma$ is the standard deviation of metric across all stations, and $N$ is the number of stations. The standard error reflects the uncertainty in the estimation of mean value reported in Table.~\ref{tab:evenbasedmet}. Note that while we present the mean and uncertainty in the metric in Table.~\ref{tab:evenbasedmet}, we report the metrics for all the stations in Supplementary Information. 

%------------------
\begin{table*}
\centering
\caption{Event based metric comparison of {\model} with {\weimer} and Persistence model, summarized across all stations. }
\label{tab:evenbasedmet}
\resizebox{\textwidth}{!}{%
\begin{tabular}{cc|cccc|cccc|cccc}
    \cline{1-14}
    \multirow{2}{*}{Storm} & \multirow{2}{*}{Metric} &   \multicolumn{4}{|c|}{\model} & \multicolumn{4}{|c}{Weimer} & \multicolumn{4}{|c}{Persistence}\\ 
    \cline{3-14}
    & & 18 & 42 & 66 & 90 & 18 & 42 & 66 & 90 & 18 & 42 & 66 & 90 \\
    \cline{1-14}
    \multirow{4}{*}{2011} & POD & 0.62 $\pm$ 0.03 & 0.58 $\pm$ 0.03 & 0.42 $\pm$ 0.04 & 0.31 $\pm$ 0.04 & 0.17$\pm$0.01 & 0.09$\pm$0.01 & 0.05$\pm$0.01 & 0.01$\pm$0.01 & 0.56$\pm$0.03 & 0.53$\pm$0.02 & 0.38$\pm$0.03 & 0.23$\pm$0.03 \\ 

    &  POFD & 0.13 $\pm$ 0.01 & 0.07 $\pm$ 0.01 & 0.05 $\pm$ 0.00 & 0.02 $\pm$ 0.00 & 0.01$\pm$0.00 & 0.00$\pm$0.00 & 0.00$\pm$0.00 & 0.00$\pm$0.00 & 0.06$\pm$0.01 & 0.03$\pm$0.00 & 0.02$\pm$0.00 & 0.01$\pm$0.00 \\ 
    
    &  PC & 0.87 $\pm$ 0.01 & 0.92 $\pm$ 0.01 & 0.94 $\pm$ 0.01 & 0.97 $\pm$ 0.00 & 0.86$\pm$0.01 & 0.94$\pm$0.01 & 0.97$\pm$0.00 & 0.98$\pm$0.00 & 0.92$\pm$0.01 & 0.95$\pm$0.00 & 0.96$\pm$0.00 & 0.98$\pm$0.00 \\ 
    
    &  HSS & 0.37 $\pm$ 0.02 & 0.30 $\pm$ 0.02 & 0.22 $\pm$ 0.03 & 0.17 $\pm$ 0.03 & 0.18$\pm$0.01 & 0.12$\pm$0.01 & 0.06$\pm$0.02 & 0.02$\pm$0.01 & 0.47$\pm$0.02 & 0.46$\pm$0.02 & 0.32$\pm$0.03 & 0.19$\pm$0.02 \\ 
    \cline{1-14}
    
    \multirow{4}{*}{2015} & POD & 0.69 $\pm$ 0.02 & 0.36 $\pm$ 0.03 & 0.15 $\pm$ 0.01 & 0.06 $\pm$ 0.01 & 0.11$\pm$0.01 & 0.02$\pm$0.00 & 0.01$\pm$0.00 & 0.00$\pm$0.00 & 0.66$\pm$0.02 & 0.48$\pm$0.02 & 0.36$\pm$0.02 & 0.30$\pm$0.02 \\ 

    &  POFD & 0.57 $\pm$ 0.02 & 0.27 $\pm$ 0.02 & 0.14 $\pm$ 0.01 & 0.07 $\pm$ 0.01 & 0.04$\pm$0.00 & 0.01$\pm$0.00 & 0.00$\pm$0.00 & 0.00$\pm$0.00 & 0.21$\pm$0.02 & 0.09$\pm$0.01 & 0.05$\pm$0.00 & 0.03$\pm$0.00 \\ 
    
    &  PC & 0.66 $\pm$ 0.01 & 0.73 $\pm$ 0.02 & 0.83 $\pm$ 0.01 & 0.90 $\pm$ 0.01 & 0.68$\pm$0.02 & 0.85$\pm$0.01 & 0.92$\pm$0.01 & 0.95$\pm$0.00 & 0.85$\pm$0.01 & 0.88$\pm$0.01 & 0.92$\pm$0.01 & 0.95$\pm$0.00 \\ 
    
    &  HSS & 0.04 $\pm$ 0.01 & 0.00 $\pm$ 0.01 & -0.02 $\pm$ 0.01 & -0.03 $\pm$ 0.00 & 0.06$\pm$0.01 & 0.01$\pm$0.00 & 0.01$\pm$0.00 & -0.00$\pm$0.00 & 0.41$\pm$0.02 & 0.36$\pm$0.01 & 0.28$\pm$0.02 & 0.25$\pm$0.02 \\ 
    \cline{1-14}
\end{tabular}%
}
\end{table*}
%--------------------------------
From Table.~\ref{tab:evenbasedmet}, we first compare the performance for the 2011 storm. We see that the metrics for all the models reduce as a function of the selected threshold. First, {\model} shows a larger POD than either {\weimer} or the Persistence model, implying many of the events are detected well by {\model}. This is in-line with {\model} being able to capture the variation in peaks of the measurements well. 

Next, we find that  {\model} shows larger POFD when compared to {\weimer} or Persistence for a threshold of $18$ nT/min. However, the POFD becomes small and consistent with the benchmarks for larger thresholds. 

Third, we find that the proportion correct from {\model} are consistent with those from {\weimer}, irrespective of the threshold value chosen. Since the POD of {\model} is larger than {\weimer}, this means that there are far more non-event cases which are captured well enough by both the models. However, the persistence model has a larger PC, which again indicates some true negatives being missed out by {\model}.

Finally, {\model} HSS are larger than {\weimer}, but smaller than the persistence model. This tells us that proportion of correct forecasts by {\model} are significantly better informed than those from {\weimer}. However, the forecasting horizon contains enough auto-correlation in the {\sg} time series to give rise to a good fraction of non-random correct proportion of events. The HSS between {\model} and Persistence become consistent only at a threshold of $90$ nT/min, indicating that the large events are not very persistent, and this information from the solar wind is captured by {\model}.

Interestingly, for the 2015 storm, all of our metrics -- both for {\model} and {\weimer} are worser than the persistence model. {\model} shows better metric performance when compared to {\weimer}, and shows only marginally better (or worse) metrics  (except HSS) when compared to persistence. However, the HSS indicates that both {\model} and {\weimer} are no better than a random model generating 1s and 0s, which is consistent with the performance of similar RNN based models~\cite{Keese_2020_GICpaper}.

Since both {\model} and {\weimer} do not give as good a performance as the persistence model, we get further evidence of the strong influence of missing OMNI data in giving rise to the poor performance of OMNI-based forecasting schemes. 

% -----------------------------------
% \begin{table}[]
%     % \centering
%     \caption{Performance of all stations for the 2011 storm, and comparison with the empirical fitting scheme of {\weimer}.}
%     \label{tab:storm2011full}
%     \resizebox{\textwidth}{!}{
%   \begin{tabular}{c|cc|cc|cccc|cccc|cccc|cccc|cc|cc|cccc|cccc|cccc|cccc|cc|cc|cccc|cccc|cccc|cccc}
%   \cline{1-61}
%     \input{storm_2011.txt}
%     \cline{1-53}
%     \end{tabular}}
% \end{table}

% \begin{table}[]
%     % \centering
%     \caption{Performance of all stations for the 2015 storm, and comparison with the empirical fitting scheme of {\weimer}.}
%     \label{tab:storm2015full}
%     \resizebox{\textwidth}{!}{
%   \begin{tabular}{c|cc|cc|cccc|cccc|cccc|cccc|cc|cc|cccc|cccc|cccc|cccc|cc|cc|cccc|cccc|cccc|cccc}
%   \cline{1-61}
%     \input{storm_2015.txt}
%     \cline{1-53}
%     \end{tabular}}
% \end{table}

%--------------------
\section{Summary and Conclusion}\label{sec:summaryconclusion}

Accurate global forecasts of geomagnetic perturbations are extremely important from the perspective of both disruptions due to GICs, and to understand the modulation Earth's global magnetic field due the streaming solar wind. 

To this end, we develop a global magnetic field perturbation forecasting model in this work. The model, named {\model}, has three components: a time series summarizer, a coefficient generator, and a spherical harmonic constructor. The time series summarizer takes in a time series of solar wind, IMF and solar radio flux time, and generates a summary state across all variables and time. This summary state is transformed non linearly by a fully connected layer to generate a vector of coefficients. Finally, the spherical harmonic layer contracts with this coefficient layer, and generates perturbation forecasts at different locations on the Earth.

We find that the {\model} is able to clearly capture the temporal variations of the perturbations. However, it under-predicts the perturbation values if they are $\approx1000$ nT, resulting in large pointwise errors. Note that {\model} is trained predominantly during quiet times -- since 2.1\% of data consists of a SYM-H index of less than -50 nT. Thus, the results may potentially be biased toward more quiet time than active times, resulting in under-prediction of perturbations.

We benchmark our model against {\weimer} and a 30-minute persistence model using various metrics. We find that {\model} clearly outperforms {\weimer} in all metrics. However, {\model} shows comparable (or slightly worse) performance than a persistence model on MAE \& RMSE. On the event based metrics, {\model} shows either consistent, or worse performance than the  persistence model. \response{Clearly, a persistence model seems to possess an advantage over both {\model} and {\weimer}. However, the persistence can be computed only for individual stations, and lacks the spatial coverage which {\model} provides.}

\response{Regardless of the exact performance measure, our results show}  that deep learning is able to capture associations between changes in the solar wind/interplanetary medium, and the Earth's magnetosphere. However, the magnetosphere seems to have enough memory over 30 minutes for a persistence model to show good enough performance. For the 2015 storm, {\model} and {\weimer} show much worse performance than persistence model -- this is not the case for the 2011 storm, where the performance is comparable. Since both {\model} and {\weimer} show the drastic reduction in performance on the 2015 storm, the underlying reason seems to be the missing values in the OMNI dataset, resulting in a noisy input to the model. However, the similarly in metrics for both the 2011 and 2015 storm for the Persistence model tells us that {\model} can, in principle, perform well given good data.

\response{Also note that the time scale of 120 minutes is interestingly of the order of time to transfer information from dayside and nightside reconnection sites in the magnetosphere to the ionosphere system especially for higher latitude $\ge 40^{\circ}$ \cite{Coxon-timelag}. However, we may only speculate, and not claim an exact connection at this stage. }

Our results may be compared across literature with models which benchmark on the two storms, at similar cadence. Models by \citeA{Keese_2020_GICpaper} consider the solar wind and IMF parameters as inputs, and output the geomagnetic perturbations at the Ottawa station. However, note that while {\model} has a forecast horizon of 30 minutes (over and above the lag between OMNI and {\sg}), such a lag is not present in \citeA{Keese_2020_GICpaper}. 

We can first compare the average metrics from {\model} with the metrics provided by \citeA{Keese_2020_GICpaper} (compare Table.~\ref{tab:evenbasedmet} of this paper with Table. 1 of \citeA{Keese_2020_GICpaper}). For the 2011 storm, {\model} clearly outperforms the LSTM model of \citeA{Keese_2020_GICpaper} in all metrics except POFD. Similarly, {\model} shows better performance than the ANN model for POD and PC, while the performance is consistent in HSS and slightly worse in POFD. For the 2015 storm, {\model} shows better performance in POD, marginally worse performance in PC and HSS, while far many false detections are made by {\model} for a threshold of 18 nT/min. This clearly seems to be a manifestation of the missing data and the imputation scheme deployed to tackle it. Hence, prima-facie, it seems that linear interpolation is much better imputation scheme than zeroing of inputs. Note, however, that {\model} forecasts are not confined to any particular station, and generates maps of forecasts.

Next, we may also pick out the specific metrics for the OTT station (presented in Table.~\ref{tab:ott}), and compare them with the two models of \citeA{Keese_2020_GICpaper}. Here, we find for the 2011 storm that while {\model} shows better performance than both the models of \citeA{Keese_2020_GICpaper} in POD and PC, while the performance is marginally worse in POFD and HSS. However, note that {\model} HSS is more than (or even similar) to the LSTM model of \citeA{Keese_2020_GICpaper}, while it is lower than the ANN model. For the 2015 storm, {\model} outperforms both the models in POD, while the performance is marginally worse in PC, POFD and HSS. This is consistent with the average performance across all stations, and seems to again point towards a dependence on the data imputation scheme. 

%-----------------------------------
\begin{table}[ht!]
\centering
\caption{Metric comparison between {\model}, {\weimer} and Persistence models for the OTT station for the years 2011 (top row) and 2015 (bottom row). Blanks (-) denote metrics which are unavailable due to the denominator in the metric definition going to 0. The full table for all stations is given in Supplementary Information.}
\label{tab:ott}
\begin{tabular}{cc|cc|cc|cc}
    \hline
     \multicolumn{2}{c|}{Metric} & \multicolumn{2}{|c}{\model} & \multicolumn{2}{|c}{\weimer}  & \multicolumn{2}{|c}{Persistence}\\
     \cline{3-8}
     & & 2011 & 2015& 2011 & 2015& 2011 & 2015 \\
     \cline{1-8}
    \multicolumn{1}{c|}{\multirow{2}{*}{MSE}}&$\delta b_e$ &22.54& 46.35 & 28.48&45.97&24.58&40.64 \\
    \multicolumn{1}{c|}{ }&  $\delta b_n$ & 52.87&79.89 &46.37&65.23&37.78&46.14\\
    \cline{2-8}
    \multicolumn{1}{c|}{\multirow{2}{*}{MAE}}&$\delta b_e$ &  14.68 &34.18  &22.51 &32.33&14.63&26.12\\
    \multicolumn{1}{c|}{} &  $\delta b_n$ & 32.55 &57.68  &35.40&54.30&20.13&27.50\\
     \cline{2-8}
    \multicolumn{1}{c|}{\multirow{4}{*}{18}}&POD& 0.56 &0.52&0.33&0.04&0.78 &0.64\\
    \multicolumn{1}{c|}{} &  POFD & 0.14 &0.42  &0.00&0.04&0.03 &0.08\\
    \multicolumn{1}{c|}{} &  PC & 0.84 &0.57 &0.95&0.82&0.95 &0.88\\
    \multicolumn{1}{c|}{} &  HSS & 0.25 &0.06 &0.48&-0.00 &0.68 &0.55\\
    \cline{2-8}
    \multicolumn{1}{c|}{\multirow{4}{*}{42}} &POD & 0.00 &0.00&0.00&0.00&0.00 &0.29\\
    \multicolumn{1}{c|}{} &  POFD & 0.00 &0.07 &0.00&0.01&0.01 &0.04\\
    \multicolumn{1}{c|}{} &  PC &0.99 & 0.89 &0.99&0.95 &0.98 &0.93\\
    \multicolumn{1}{c|}{} &  HSS & 0.00 &-0.06 &0.00&-0.01 &-0.01 &0.21\\
    \cline{2-8}
    \multicolumn{1}{c|}{\multirow{4}{*}{66}}& POD & - &0.00&-&0.00&- &0.50\\
    \multicolumn{1}{c|}{} &  POFD & 0.00 &0.01 &0.00&0.00&0.00 &0.01\\
    \multicolumn{1}{c|}{} &  PC & 1.00 &0.98 &1.00&0.99&1.00 &0.99\\
    \multicolumn{1}{c|}{} &  HSS & - &-0.01  &- &0.00 &- &0.49\\
     \cline{2-8}
    \multicolumn{1}{c|}{\multirow{4}{*}{90}}& POD &- & 0.00 &-&0.00 &- &0.00\\
    \multicolumn{1}{c|}{} &  POFD & 0.00 &0.01 &0.00&0.00&0.00&0.01\\
    \multicolumn{1}{c|}{} &  PC & 1.00 &0.99&1.00&0.99&1.00&0.98\\
    \multicolumn{1}{c|}{} &  HSS & - &-0.01 &-&0.00&-&-0.01\\
    \hline 
\end{tabular}
\end{table}
%-----------------------------------

We also compare our result with \citeA{Pulkkinen_metrics} study but not for a particular station or event. In general, none of the models in the community, including first-principle and empirical, can capture the high dB/dt (1.5 nT/s or 90 nT/min) threshold. \response{This behaviour is very important while forecasting particularly strong spaceweather events. If the mitigation of a storm depends on a model forecast, under-prediction of the perturbation magnitude would pose a significant problem. These models are not able to reproduce point-by-point fluctuations of perturbation due to the complex waveform of the perturbation signal. Hence, this is an important issue which would need to be mitigated in the future.} 

It is important to note the various caveats associated with this work. The first, and the most obvious issue is of the missing data. We have imputed the missing data with $0$s. While this is a simplistic scheme of imputation, we did not perform any interpolation as we did not find any well-motivated reason to induce artificial variations in the data. However, addressing this issue with complex imputation schemes is far too complicated, and beyond the scope of this work. 

Next, we see that the {\model} forecasts follow the variations in {\sg} measurements well, but do not reproduce the exact values when the perturbations are large. The fact that {\model} captures the variations, but not the exact magnitude, seems to arise from a lack of ``context'' perturbation measurements. \response{One can perform a non-linear rescaling \cite<see for example.>{Enrico_2020_NonlinearRescaling} to circumvent this issue. However, our overarching aim is to have a rather more self-consistent model avoiding any ad-hoc scaling as much as possible.} In principle, we can incorporate a proxy for the state of the Earth's magnetosphere as an input to the model. This would help provide ``context'' to the forecasting model, and may help it give the correct perturbation values (and not just capture the variations). \response{Incorporating geomagnetic indices have been shown to improve the quality of magentospheric forecasts\cite<see, for ex.>{2020_Smith_includeindices}. However,} capturing a summary of the magnetosphere given changing stations across multiple MLT and MAGLAT is non trivial, and is a work for the future. 

Third, our method assumes a smooth, continuous and differentiable perturbation field, with power distributed amongst different modes. Furthermore, we truncate the spherical harmonics at a maximum mode due to operational constraints and hyperparameter selection. While these assumptions are physically motivated, their effect is to impose a ``smooth'' reconstruction, which may prevent capture of localized large peaks in data across a set of (MLT,MAGLAT).\response{Similarly, since we have truncated the spherical harmonics at a maximum number of modes, we expect the highest frequency mode to be translated to the shortest length scale that our workflow can resolve. Hence, {\model} will not be able to -- in the current formulation -- resolve fluctuations shorter than this ``threshold'' length scale. Note further that the shortest length scales of importance would also depend on local ionospheric current and local geology. We however expect these scales to however be much smaller than the length scale corresponding to the highest harmonic mode considered~\cite{beggan2015sensitivity,2015_Pulkinnen_lengthscale}.}

\response{The spherical harmonic formulation performs an instantaneous decomposition of field over the globe. However,} the whole system -- as a sphere -- evolves dynamically over time. \response{Hence}, information propagation across different stations takes time, which must be incorporated in the basis matrix formulation itself. While this is beyond the scope of the paper, such a path is a potential future work for improvement. 

Also, note that {\model} does not yet provide uncertainty estimates on the perturbation forecasts. The uncertainty estimates both provide a degree of confidence, and also inform us of ill-constrained regions of forecasts. Thus, such uncertainties may provide us with means of diagnosing the most optimum location of stations to (i). reduce uncertainty, and (ii). optimize the number of stations.

Finally, we emphasize that the codebase and the proposed model {\model} are general enough to be suited for forecasting fields on any spherically-symmetric systems. A direct application of {\model} would be transfer-learn the perturbation forecasts to magnetic field perturbation measurements in other planets. This is useful from both a spacecraft navigation, and a science measurement perspective to gather data pertaining to specific locations as a study of planetary magnetospheres.

%-----------
\section{Open Research}
Our model outputs are agnostic to the grid on which the basis is defined. Hence, the coefficients may be contracted with an appropriate basis to generate full-Earth maps of perturbation forecasts. To further reproducible research, and foster innovation with modelling schemes, we are making our codebase and models open source at \citeA{Daggermodel_github}. For getting researchers started with using our code, a tutorial notebook as a part of SpaceML~\cite{spaceml} is available at \url{https://spaceml.org/repo/project/60c0a78d4ba8cb0012611ad4}.

Our model is built in PyTorch~\cite{pytorch}, PyTorch-lightning~\cite{pytorchlightning} and Sympy~\cite{sympy}. We also use Numpy~\cite{numpy}, Scikit-learn~\cite{scikit-learn} and  Scipy~\cite{scipy} for analysis, Dask~\cite{dask} and Pandas~\cite{pandas} for data processing, and Matplotlib~\cite{matplotlib}, Cartopy~\cite{Cartopy} for plotting.
%-----------------------
\acknowledgments
This research was conducted as a part of research sprint at Frontier Development Lab (FDL). We would like to thank SETI, Google Cloud, Nvidia and Lockheed Martin for funding the FDL program. We thank the SuperMAG collaboration for making the perturbation data available for the public. We also sincerely thank Prof. Daniel Weimer for providing the storm time forecasts, which served as excellent benchmarks. We thank the two reviewers for their very useful comments. U.V would like to acknowledge the Max Planck Partner Grant of Prof. Durgesh Tripathi, IUCAA for providing compute facility, and the Nvidia Academic Hardware program grant to U.V \& Durgesh Tripathi. B.F was supported by NASA grant NNH19ZDA001N-LWS. P.T. is supported by the UK EPSRC CDT in Autonomous Intelligent Machines and Systems (grant reference EP/L015897/1.
%-------------------------
\bibliography{geoeffectivness.bib}

\begin{thebibliography}{}

\bibitem [\protect \citeauthoryear {%
Barlow%
, Barlow%
\BCBL {}\ \BBA {} Culley%
}{%
Barlow%
\ \protect \BOthers {.}}{%
{\protect \APACyear {1849}}%
}]{%
barlow1849}
\APACinsertmetastar {%
barlow1849}%
\begin{APACrefauthors}%
Barlow, W\BPBI H.%
, Barlow, P.%
\BCBL {}\ \BBA {} Culley, R\BPBI S.%
\end{APACrefauthors}%
\unskip\
\newblock
\APACrefYearMonthDay{1849}{}{}.
\newblock
{\BBOQ}\APACrefatitle {VI. On the spontaneous electrical currents observed in
  the wires of the electric telegraph} {Vi. on the spontaneous electrical
  currents observed in the wires of the electric telegraph}.{\BBCQ}
\newblock
\APACjournalVolNumPages{Philosophical Transactions of the Royal Society of
  London}{139}{}{61-72}.
\newblock
\begin{APACrefURL}
  \url{https://royalsocietypublishing.org/doi/abs/10.1098/rstl.1849.0006}
  \end{APACrefURL}
\newblock
\begin{APACrefDOI} \doi{10.1098/rstl.1849.0006} \end{APACrefDOI}
\PrintBackRefs{\CurrentBib}

\bibitem [\protect \citeauthoryear {%
Beggan%
}{%
Beggan%
}{%
{\protect \APACyear {2015}}%
}]{%
beggan2015sensitivity}
\APACinsertmetastar {%
beggan2015sensitivity}%
\begin{APACrefauthors}%
Beggan, C\BPBI D.%
\end{APACrefauthors}%
\unskip\
\newblock
\APACrefYearMonthDay{2015}{}{}.
\newblock
{\BBOQ}\APACrefatitle {Sensitivity of geomagnetically induced currents to
  varying auroral electrojet and conductivity models} {Sensitivity of
  geomagnetically induced currents to varying auroral electrojet and
  conductivity models}.{\BBCQ}
\newblock
\APACjournalVolNumPages{Earth, Planets and Space}{67}{1}{1--12}.
\PrintBackRefs{\CurrentBib}

\bibitem [\protect \citeauthoryear {%
Biewald%
}{%
Biewald%
}{%
{\protect \APACyear {2020}}%
}]{%
wandb}
\APACinsertmetastar {%
wandb}%
\begin{APACrefauthors}%
Biewald, L.%
\end{APACrefauthors}%
\unskip\
\newblock
\APACrefYearMonthDay{2020}{}{}.
\newblock
\APACrefbtitle {Experiment Tracking with Weights and Biases.} {Experiment
  tracking with weights and biases.}
\newblock
\begin{APACrefURL} \url{https://www.wandb.com/} \end{APACrefURL}
\newblock
\APACrefnote{Software available from wandb.com}
\PrintBackRefs{\CurrentBib}

\bibitem [\protect \citeauthoryear {%
{Boteler}%
}{%
{Boteler}%
}{%
{\protect \APACyear {2001}}%
}]{%
boteler-2001}
\APACinsertmetastar {%
boteler-2001}%
\begin{APACrefauthors}%
{Boteler}, D\BPBI H.%
\end{APACrefauthors}%
\unskip\
\newblock
\APACrefYearMonthDay{2001}{{\APACmonth{01}}}{}.
\newblock
{\BBOQ}\APACrefatitle {{Space weather effects on power systems}} {{Space
  weather effects on power systems}}.{\BBCQ}
\newblock
\APACjournalVolNumPages{Washington DC American Geophysical Union Geophysical
  Monograph Series}{125}{}{347-352}.
\newblock
\begin{APACrefDOI} \doi{10.1029/GM125p0347} \end{APACrefDOI}
\PrintBackRefs{\CurrentBib}

\bibitem [\protect \citeauthoryear {%
Camporeale%
}{%
Camporeale%
}{%
{\protect \APACyear {2019}}%
}]{%
camporeale-2019}
\APACinsertmetastar {%
camporeale-2019}%
\begin{APACrefauthors}%
Camporeale, E.%
\end{APACrefauthors}%
\unskip\
\newblock
\APACrefYearMonthDay{2019}{{\APACmonth{08}}}{}.
\newblock
{\BBOQ}\APACrefatitle {The {{Challenge}} of {{Machine Learning}} in {{Space
  Weather Nowcasting}} and {{Forecasting}}} {The {{Challenge}} of {{Machine
  Learning}} in {{Space Weather Nowcasting}} and {{Forecasting}}}.{\BBCQ}
\newblock
\APACjournalVolNumPages{Space Weather}{17}{8}{1166--1207}.
\newblock
\begin{APACrefDOI} \doi{10.1029/2018SW002061} \end{APACrefDOI}
\PrintBackRefs{\CurrentBib}

\bibitem [\protect \citeauthoryear {%
{Camporeale}%
\ \protect \BOthers {.}}{%
{Camporeale}%
\ \protect \BOthers {.}}{%
{\protect \APACyear {2020}}%
}]{%
Enrico_2020_NonlinearRescaling}
\APACinsertmetastar {%
Enrico_2020_NonlinearRescaling}%
\begin{APACrefauthors}%
{Camporeale}, E.%
, {Cash}, M\BPBI D.%
, {Singer}, H\BPBI J.%
, {Balch}, C\BPBI C.%
, {Huang}, Z.%
\BCBL {}\ \BBA {} {Toth}, G.%
\end{APACrefauthors}%
\unskip\
\newblock
\APACrefYearMonthDay{2020}{{\APACmonth{11}}}{}.
\newblock
{\BBOQ}\APACrefatitle {{A Gray-Box Model for a Probabilistic Estimate of
  Regional Ground Magnetic Perturbations: Enhancing the NOAA Operational
  Geospace Model With Machine Learning}} {{A Gray-Box Model for a Probabilistic
  Estimate of Regional Ground Magnetic Perturbations: Enhancing the NOAA
  Operational Geospace Model With Machine Learning}}.{\BBCQ}
\newblock
\APACjournalVolNumPages{Journal of Geophysical Research (Space
  Physics)}{125}{11}{e27684}.
\newblock
\begin{APACrefDOI} \doi{10.1029/2019JA027684} \end{APACrefDOI}
\PrintBackRefs{\CurrentBib}

\bibitem [\protect \citeauthoryear {%
Cho%
\ \protect \BOthers {.}}{%
Cho%
\ \protect \BOthers {.}}{%
{\protect \APACyear {2014}}%
}]{%
cho2014learning}
\APACinsertmetastar {%
cho2014learning}%
\begin{APACrefauthors}%
Cho, K.%
, Van~Merri{\"e}nboer, B.%
, Gulcehre, C.%
, Bahdanau, D.%
, Bougares, F.%
, Schwenk, H.%
\BCBL {}\ \BBA {} Bengio, Y.%
\end{APACrefauthors}%
\unskip\
\newblock
\APACrefYearMonthDay{2014}{}{}.
\newblock
{\BBOQ}\APACrefatitle {Learning Phrase Representations Using {{RNN}}
  Encoder-Decoder for Statistical Machine Translation} {Learning phrase
  representations using {{RNN}} encoder-decoder for statistical machine
  translation}.{\BBCQ}
\newblock
\APACjournalVolNumPages{arXiv preprint arXiv:1406.1078}{}{}{}.
\PrintBackRefs{\CurrentBib}

\bibitem [\protect \citeauthoryear {%
{Clette}%
}{%
{Clette}%
}{%
{\protect \APACyear {2021}}%
}]{%
F107_Frederic}
\APACinsertmetastar {%
F107_Frederic}%
\begin{APACrefauthors}%
{Clette}, F.%
\end{APACrefauthors}%
\unskip\
\newblock
\APACrefYearMonthDay{2021}{{\APACmonth{01}}}{}.
\newblock
{\BBOQ}\APACrefatitle {{Is the F$_{10.7cm}$ - Sunspot Number relation linear
  and stable?}} {{Is the F$_{10.7cm}$ - Sunspot Number relation linear and
  stable?}}{\BBCQ}
\newblock
\APACjournalVolNumPages{Journal of Space Weather and Space Climate}{11}{}{2}.
\newblock
\begin{APACrefDOI} \doi{10.1051/swsc/2020071} \end{APACrefDOI}
\PrintBackRefs{\CurrentBib}

\bibitem [\protect \citeauthoryear {%
Coxon%
\ \protect \BOthers {.}}{%
Coxon%
\ \protect \BOthers {.}}{%
{\protect \APACyear {2019}}%
}]{%
Coxon-timelag}
\APACinsertmetastar {%
Coxon-timelag}%
\begin{APACrefauthors}%
Coxon, J\BPBI C.%
, Shore, R\BPBI M.%
, Freeman, M\BPBI P.%
, Fear, R\BPBI C.%
, Browett, S\BPBI D.%
, Smith, A\BPBI W.%
\BDBL {}Anderson, B\BPBI J.%
\end{APACrefauthors}%
\unskip\
\newblock
\APACrefYearMonthDay{2019}{jul}{}.
\newblock
{\BBOQ}\APACrefatitle {Timescales of Birkeland Currents Driven by the IMF}
  {Timescales of birkeland currents driven by the imf}.{\BBCQ}
\newblock
\APACjournalVolNumPages{Geophysical Research Letters}{46}{}{7893-7901}.
\newblock
\begin{APACrefURL} \url{https://oadoi.org/10.1029/2018gl081658}
  \end{APACrefURL}
\newblock
\begin{APACrefDOI} \doi{10.1029/2018gl081658} \end{APACrefDOI}
\PrintBackRefs{\CurrentBib}

\bibitem [\protect \citeauthoryear {%
{Dask Development Team}%
}{%
{Dask Development Team}%
}{%
{\protect \APACyear {2016}}%
}]{%
dask}
\APACinsertmetastar {%
dask}%
\begin{APACrefauthors}%
{Dask Development Team}.%
\end{APACrefauthors}%
\unskip\
\newblock
\APACrefYearMonthDay{2016}{}{}.
\newblock
{\BBOQ}\APACrefatitle {Dask: Library for dynamic task scheduling} {Dask:
  Library for dynamic task scheduling}{\BBCQ}\ [\bibcomputersoftwaremanual].
\newblock
\begin{APACrefURL} \url{https://dask.org} \end{APACrefURL}
\PrintBackRefs{\CurrentBib}

\bibitem [\protect \citeauthoryear {%
development team%
}{%
development team%
}{%
{\protect \APACyear {2020}}%
}]{%
pandas}
\APACinsertmetastar {%
pandas}%
\begin{APACrefauthors}%
development team, T\BPBI P.%
\end{APACrefauthors}%
\unskip\
\newblock
\APACrefYearMonthDay{2020}{{\APACmonth{02}}}{}.
\newblock
\APACrefbtitle {pandas-dev/pandas: Pandas.} {pandas-dev/pandas: Pandas.}
\newblock
\APACaddressPublisher{}{Zenodo}.
\newblock
\begin{APACrefURL} \url{https://doi.org/10.5281/zenodo.3509134}
  \end{APACrefURL}
\newblock
\begin{APACrefDOI} \doi{10.5281/zenodo.3509134} \end{APACrefDOI}
\PrintBackRefs{\CurrentBib}

\bibitem [\protect \citeauthoryear {%
Eastwood%
\ \protect \BOthers {.}}{%
Eastwood%
\ \protect \BOthers {.}}{%
{\protect \APACyear {2018}}%
}]{%
eastwood2018quantifying}
\APACinsertmetastar {%
eastwood2018quantifying}%
\begin{APACrefauthors}%
Eastwood, J.%
, Hapgood, M.%
, Biffis, E.%
, Benedetti, D.%
, Bisi, M.%
, Green, L.%
\BDBL {}Burnett, C.%
\end{APACrefauthors}%
\unskip\
\newblock
\APACrefYearMonthDay{2018}{}{}.
\newblock
{\BBOQ}\APACrefatitle {Quantifying the economic value of space weather
  forecasting for power grids: An exploratory study} {Quantifying the economic
  value of space weather forecasting for power grids: An exploratory
  study}.{\BBCQ}
\newblock
\APACjournalVolNumPages{Space weather}{16}{12}{2052--2067}.
\PrintBackRefs{\CurrentBib}

\bibitem [\protect \citeauthoryear {%
{Falcon et al.}%
}{%
{Falcon et al.}%
}{%
{\protect \APACyear {2019}}%
}]{%
pytorchlightning}
\APACinsertmetastar {%
pytorchlightning}%
\begin{APACrefauthors}%
{Falcon et al.}, W.%
\end{APACrefauthors}%
\unskip\
\newblock
\APACrefYearMonthDay{2019}{}{}.
\newblock
{\BBOQ}\APACrefatitle {PyTorch Lightning} {Pytorch lightning}.{\BBCQ}
\newblock
\APACjournalVolNumPages{GitHub. Note:
  https://github.com/PyTorchLightning/pytorch-lightning}{3}{}{}.
\PrintBackRefs{\CurrentBib}

\bibitem [\protect \citeauthoryear {%
Gjerloev%
}{%
Gjerloev%
}{%
{\protect \APACyear {2012}}%
}]{%
gjerloev2012supermag}
\APACinsertmetastar {%
gjerloev2012supermag}%
\begin{APACrefauthors}%
Gjerloev, J.%
\end{APACrefauthors}%
\unskip\
\newblock
\APACrefYearMonthDay{2012}{}{}.
\newblock
{\BBOQ}\APACrefatitle {The SuperMAG data processing technique} {The supermag
  data processing technique}.{\BBCQ}
\newblock
\APACjournalVolNumPages{Journal of Geophysical Research: Space
  Physics}{117}{A9}{}.
\PrintBackRefs{\CurrentBib}

\bibitem [\protect \citeauthoryear {%
Harris%
\ \protect \BOthers {.}}{%
Harris%
\ \protect \BOthers {.}}{%
{\protect \APACyear {2020}}%
}]{%
numpy}
\APACinsertmetastar {%
numpy}%
\begin{APACrefauthors}%
Harris, C\BPBI R.%
, Millman, K\BPBI J.%
, van~der Walt, S\BPBI J.%
, Gommers, R.%
, Virtanen, P.%
, Cournapeau, D.%
\BDBL {}Oliphant, T\BPBI E.%
\end{APACrefauthors}%
\unskip\
\newblock
\APACrefYearMonthDay{2020}{{\APACmonth{09}}}{}.
\newblock
{\BBOQ}\APACrefatitle {Array programming with {NumPy}} {Array programming with
  {NumPy}}.{\BBCQ}
\newblock
\APACjournalVolNumPages{Nature}{585}{7825}{357--362}.
\newblock
\begin{APACrefURL} \url{https://doi.org/10.1038/s41586-020-2649-2}
  \end{APACrefURL}
\newblock
\begin{APACrefDOI} \doi{10.1038/s41586-020-2649-2} \end{APACrefDOI}
\PrintBackRefs{\CurrentBib}

\bibitem [\protect \citeauthoryear {%
Hochreiter%
\ \BBA {} Schmidhuber%
}{%
Hochreiter%
\ \BBA {} Schmidhuber%
}{%
{\protect \APACyear {1997}}%
}]{%
hochreiter1997long}
\APACinsertmetastar {%
hochreiter1997long}%
\begin{APACrefauthors}%
Hochreiter, S.%
\BCBT {}\ \BBA {} Schmidhuber, J.%
\end{APACrefauthors}%
\unskip\
\newblock
\APACrefYearMonthDay{1997}{}{}.
\newblock
{\BBOQ}\APACrefatitle {Long short-term memory} {Long short-term memory}.{\BBCQ}
\newblock
\APACjournalVolNumPages{Neural computation}{9}{8}{1735--1780}.
\PrintBackRefs{\CurrentBib}

\bibitem [\protect \citeauthoryear {%
Hunter%
}{%
Hunter%
}{%
{\protect \APACyear {2007}}%
}]{%
matplotlib}
\APACinsertmetastar {%
matplotlib}%
\begin{APACrefauthors}%
Hunter, J\BPBI D.%
\end{APACrefauthors}%
\unskip\
\newblock
\APACrefYearMonthDay{2007}{}{}.
\newblock
{\BBOQ}\APACrefatitle {Matplotlib: A 2D graphics environment} {Matplotlib: A 2d
  graphics environment}.{\BBCQ}
\newblock
\APACjournalVolNumPages{Computing in Science \& Engineering}{9}{3}{90--95}.
\newblock
\begin{APACrefDOI} \doi{10.1109/MCSE.2007.55} \end{APACrefDOI}
\PrintBackRefs{\CurrentBib}

\bibitem [\protect \citeauthoryear {%
Keesee%
\ \protect \BOthers {.}}{%
Keesee%
\ \protect \BOthers {.}}{%
{\protect \APACyear {2020}}%
}]{%
Keese_2020_GICpaper}
\APACinsertmetastar {%
Keese_2020_GICpaper}%
\begin{APACrefauthors}%
Keesee, A\BPBI M.%
, Pinto, V.%
, Coughlan, M.%
, Lennox, C.%
, Mahmud, M\BPBI S.%
\BCBL {}\ \BBA {} Connor, H\BPBI K.%
\end{APACrefauthors}%
\unskip\
\newblock
\APACrefYearMonthDay{2020}{}{}.
\newblock
{\BBOQ}\APACrefatitle {Comparison of Deep Learning Techniques to Model
  Connections Between Solar Wind and Ground Magnetic Perturbations} {Comparison
  of deep learning techniques to model connections between solar wind and
  ground magnetic perturbations}.{\BBCQ}
\newblock
\APACjournalVolNumPages{Frontiers in Astronomy and Space Sciences}{7}{}{72}.
\newblock
\begin{APACrefURL}
  \url{https://www.frontiersin.org/article/10.3389/fspas.2020.550874}
  \end{APACrefURL}
\newblock
\begin{APACrefDOI} \doi{10.3389/fspas.2020.550874} \end{APACrefDOI}
\PrintBackRefs{\CurrentBib}

\bibitem [\protect \citeauthoryear {%
King%
\ \BBA {} Papitashvili%
}{%
King%
\ \BBA {} Papitashvili%
}{%
{\protect \APACyear {2005}}%
}]{%
king_2005_omni}
\APACinsertmetastar {%
king_2005_omni}%
\begin{APACrefauthors}%
King, J\BPBI H.%
\BCBT {}\ \BBA {} Papitashvili, N\BPBI E.%
\end{APACrefauthors}%
\unskip\
\newblock
\APACrefYearMonthDay{2005}{}{}.
\newblock
{\BBOQ}\APACrefatitle {Solar wind spatial scales in and comparisons of hourly
  Wind and ACE plasma and magnetic field data} {Solar wind spatial scales in
  and comparisons of hourly wind and ace plasma and magnetic field
  data}.{\BBCQ}
\newblock
\APACjournalVolNumPages{Journal of Geophysical Research: Space
  Physics}{110}{A2}{}.
\newblock
\begin{APACrefURL}
  \url{https://agupubs.onlinelibrary.wiley.com/doi/abs/10.1029/2004JA010649}
  \end{APACrefURL}
\newblock
\begin{APACrefDOI} \doi{https://doi.org/10.1029/2004JA010649} \end{APACrefDOI}
\PrintBackRefs{\CurrentBib}

\bibitem [\protect \citeauthoryear {%
{Koul}%
, {Ganju}%
, {Kasam}%
\BCBL {}\ \BBA {} {Parr}%
}{%
{Koul}%
\ \protect \BOthers {.}}{%
{\protect \APACyear {2020}}%
}]{%
spaceml}
\APACinsertmetastar {%
spaceml}%
\begin{APACrefauthors}%
{Koul}, A.%
, {Ganju}, S.%
, {Kasam}, M.%
\BCBL {}\ \BBA {} {Parr}, J.%
\end{APACrefauthors}%
\unskip\
\newblock
\APACrefYearMonthDay{2020}{{\APACmonth{12}}}{}.
\newblock
{\BBOQ}\APACrefatitle {{SpaceML: Distributed Open-source Research with Citizen
  Scientists for the Advancement of Space Technology for NASA}} {{SpaceML:
  Distributed Open-source Research with Citizen Scientists for the Advancement
  of Space Technology for NASA}}.{\BBCQ}
\newblock
\APACjournalVolNumPages{arXiv e-prints}{}{}{arXiv:2012.10610}.
\PrintBackRefs{\CurrentBib}

\bibitem [\protect \citeauthoryear {%
Kozyreva%
, Pilipenko%
, Belakhovsky%
\BCBL {}\ \BBA {} Sakharov%
}{%
Kozyreva%
\ \protect \BOthers {.}}{%
{\protect \APACyear {2018}}%
}]{%
gic_kozyreva2018ground}
\APACinsertmetastar {%
gic_kozyreva2018ground}%
\begin{APACrefauthors}%
Kozyreva, O\BPBI V.%
, Pilipenko, V\BPBI A.%
, Belakhovsky, V\BPBI B.%
\BCBL {}\ \BBA {} Sakharov, Y\BPBI A.%
\end{APACrefauthors}%
\unskip\
\newblock
\APACrefYearMonthDay{2018}{}{}.
\newblock
{\BBOQ}\APACrefatitle {Ground geomagnetic field and GIC response to March 17,
  2015, storm} {Ground geomagnetic field and gic response to march 17, 2015,
  storm}.{\BBCQ}
\newblock
\APACjournalVolNumPages{Earth, Planets and Space}{70}{1}{1--13}.
\PrintBackRefs{\CurrentBib}

\bibitem [\protect \citeauthoryear {%
{Lamb}%
\ \protect \BOthers {.}}{%
{Lamb}%
\ \protect \BOthers {.}}{%
{\protect \APACyear {2019}}%
}]{%
lamb_2019_gnss}
\APACinsertmetastar {%
lamb_2019_gnss}%
\begin{APACrefauthors}%
{Lamb}, K.%
, {Malhotra}, G.%
, {Vlontzos}, A.%
, {Wagstaff}, E.%
, {G{\"u}nes Baydin}, A.%
, {Bhiwandiwalla}, A.%
\BDBL {}{Bhatt}, A.%
\end{APACrefauthors}%
\unskip\
\newblock
\APACrefYearMonthDay{2019}{{\APACmonth{10}}}{}.
\newblock
{\BBOQ}\APACrefatitle {{Correlation of Auroral Dynamics and GNSS Scintillation
  with an Autoencoder}} {{Correlation of Auroral Dynamics and GNSS
  Scintillation with an Autoencoder}}.{\BBCQ}
\newblock
\APACjournalVolNumPages{arXiv e-prints}{}{}{arXiv:1910.03085}.
\PrintBackRefs{\CurrentBib}

\bibitem [\protect \citeauthoryear {%
Lanzerotti%
}{%
Lanzerotti%
}{%
{\protect \APACyear {2001}}%
}]{%
gic_lanzerotti2001space}
\APACinsertmetastar {%
gic_lanzerotti2001space}%
\begin{APACrefauthors}%
Lanzerotti, L\BPBI J.%
\end{APACrefauthors}%
\unskip\
\newblock
\APACrefYearMonthDay{2001}{}{}.
\newblock
{\BBOQ}\APACrefatitle {Space weather effects on technologies} {Space weather
  effects on technologies}.{\BBCQ}
\newblock
\APACjournalVolNumPages{Washington DC American Geophysical Union Geophysical
  Monograph Series}{125}{}{11--22}.
\PrintBackRefs{\CurrentBib}

\bibitem [\protect \citeauthoryear {%
{Met Office}%
}{%
{Met Office}%
}{%
{\protect \APACyear {2010 - 2015}}%
}]{%
Cartopy}
\APACinsertmetastar {%
Cartopy}%
\begin{APACrefauthors}%
{Met Office}.%
\end{APACrefauthors}%
\unskip\
\newblock
\APACrefYearMonthDay{2010 - 2015}{}{}.
\newblock
{\BBOQ}\APACrefatitle {Cartopy: a cartographic python library with a matplotlib
  interface} {Cartopy: a cartographic python library with a matplotlib
  interface}{\BBCQ}\ [\bibcomputersoftwaremanual].
\newblock
\APACaddressPublisher{Exeter, Devon}{}.
\newblock
\begin{APACrefURL} \url{http://scitools.org.uk/cartopy} \end{APACrefURL}
\PrintBackRefs{\CurrentBib}

\bibitem [\protect \citeauthoryear {%
Meurer%
\ \protect \BOthers {.}}{%
Meurer%
\ \protect \BOthers {.}}{%
{\protect \APACyear {2017}}%
}]{%
sympy}
\APACinsertmetastar {%
sympy}%
\begin{APACrefauthors}%
Meurer, A.%
, Smith, C\BPBI P.%
, Paprocki, M.%
, \v{C}ert\'{i}k, O.%
, Kirpichev, S\BPBI B.%
, Rocklin, M.%
\BDBL {}Scopatz, A.%
\end{APACrefauthors}%
\unskip\
\newblock
\APACrefYearMonthDay{2017}{{\APACmonth{01}}}{}.
\newblock
{\BBOQ}\APACrefatitle {SymPy: symbolic computing in Python} {Sympy: symbolic
  computing in python}.{\BBCQ}
\newblock
\APACjournalVolNumPages{PeerJ Computer Science}{3}{}{e103}.
\newblock
\begin{APACrefURL} \url{https://doi.org/10.7717/peerj-cs.103} \end{APACrefURL}
\newblock
\begin{APACrefDOI} \doi{10.7717/peerj-cs.103} \end{APACrefDOI}
\PrintBackRefs{\CurrentBib}

\bibitem [\protect \citeauthoryear {%
Ngwira%
\ \protect \BOthers {.}}{%
Ngwira%
\ \protect \BOthers {.}}{%
{\protect \APACyear {2018}}%
}]{%
gic_ngwira2018study}
\APACinsertmetastar {%
gic_ngwira2018study}%
\begin{APACrefauthors}%
Ngwira, C\BPBI M.%
, Sibeck, D.%
, Silveira, M\BPBI V.%
, Georgiou, M.%
, Weygand, J\BPBI M.%
, Nishimura, Y.%
\BCBL {}\ \BBA {} Hampton, D.%
\end{APACrefauthors}%
\unskip\
\newblock
\APACrefYearMonthDay{2018}{}{}.
\newblock
{\BBOQ}\APACrefatitle {A Study of Intense Local d B/dt Variations During Two
  Geomagnetic Storms} {A study of intense local d b/dt variations during two
  geomagnetic storms}.{\BBCQ}
\newblock
\APACjournalVolNumPages{Space Weather}{16}{6}{676--693}.
\PrintBackRefs{\CurrentBib}

\bibitem [\protect \citeauthoryear {%
{Oughton}%
, {Skelton}%
, {Horne}%
, {Thomson}%
\BCBL {}\ \BBA {} {Gaunt}%
}{%
{Oughton}%
\ \protect \BOthers {.}}{%
{\protect \APACyear {2017}}%
}]{%
Oughton:2017}
\APACinsertmetastar {%
Oughton:2017}%
\begin{APACrefauthors}%
{Oughton}, E\BPBI J.%
, {Skelton}, A.%
, {Horne}, R\BPBI B.%
, {Thomson}, A\BPBI W\BPBI P.%
\BCBL {}\ \BBA {} {Gaunt}, C\BPBI T.%
\end{APACrefauthors}%
\unskip\
\newblock
\APACrefYearMonthDay{2017}{{\APACmonth{01}}}{}.
\newblock
{\BBOQ}\APACrefatitle {{Quantifying the daily economic impact of extreme space
  weather due to failure in electricity transmission infrastructure}}
  {{Quantifying the daily economic impact of extreme space weather due to
  failure in electricity transmission infrastructure}}.{\BBCQ}
\newblock
\APACjournalVolNumPages{Space Weather}{15}{1}{65-83}.
\newblock
\begin{APACrefDOI} \doi{10.1002/2016SW001491} \end{APACrefDOI}
\PrintBackRefs{\CurrentBib}

\bibitem [\protect \citeauthoryear {%
Papitashvili%
, Bilitza%
\BCBL {}\ \BBA {} King%
}{%
Papitashvili%
\ \protect \BOthers {.}}{%
{\protect \APACyear {2014}}%
}]{%
papitashvili2014omni}
\APACinsertmetastar {%
papitashvili2014omni}%
\begin{APACrefauthors}%
Papitashvili, N.%
, Bilitza, D.%
\BCBL {}\ \BBA {} King, J.%
\end{APACrefauthors}%
\unskip\
\newblock
\APACrefYearMonthDay{2014}{}{}.
\newblock
{\BBOQ}\APACrefatitle {{{OMNI}}: {{A}} Description of near-Earth Solar Wind
  Environment} {{{OMNI}}: {{A}} description of near-earth solar wind
  environment}.{\BBCQ}
\newblock
\APACjournalVolNumPages{cosp}{40}{}{C0--1}.
\PrintBackRefs{\CurrentBib}

\bibitem [\protect \citeauthoryear {%
Paszke%
\ \protect \BOthers {.}}{%
Paszke%
\ \protect \BOthers {.}}{%
{\protect \APACyear {2019}}%
}]{%
pytorch}
\APACinsertmetastar {%
pytorch}%
\begin{APACrefauthors}%
Paszke, A.%
, Gross, S.%
, Massa, F.%
, Lerer, A.%
, Bradbury, J.%
, Chanan, G.%
\BDBL {}Chintala, S.%
\end{APACrefauthors}%
\unskip\
\newblock
\APACrefYearMonthDay{2019}{}{}.
\newblock
{\BBOQ}\APACrefatitle {PyTorch: An Imperative Style, High-Performance Deep
  Learning Library} {Pytorch: An imperative style, high-performance deep
  learning library}.{\BBCQ}
\newblock
\BIn{} H.~Wallach, H.~Larochelle, A.~Beygelzimer, F.~d\textquotesingle
  Alch\'{e}-Buc, E.~Fox\BCBL {}\ \BBA {} R.~Garnett\ (\BEDS), \APACrefbtitle
  {Advances in Neural Information Processing Systems 32} {Advances in neural
  information processing systems 32}\ (\BPGS\ 8024--8035).
\newblock
\APACaddressPublisher{}{Curran Associates, Inc.}
\newblock
\begin{APACrefURL}
  \url{http://papers.neurips.cc/paper/9015-pytorch-an-imperative-style-high-performance-deep-learning-library.pdf}
  \end{APACrefURL}
\PrintBackRefs{\CurrentBib}

\bibitem [\protect \citeauthoryear {%
Pedregosa%
\ \protect \BOthers {.}}{%
Pedregosa%
\ \protect \BOthers {.}}{%
{\protect \APACyear {2011}}%
}]{%
scikit-learn}
\APACinsertmetastar {%
scikit-learn}%
\begin{APACrefauthors}%
Pedregosa, F.%
, Varoquaux, G.%
, Gramfort, A.%
, Michel, V.%
, Thirion, B.%
, Grisel, O.%
\BDBL {}Duchesnay, E.%
\end{APACrefauthors}%
\unskip\
\newblock
\APACrefYearMonthDay{2011}{}{}.
\newblock
{\BBOQ}\APACrefatitle {Scikit-learn: Machine Learning in {P}ython}
  {Scikit-learn: Machine learning in {P}ython}.{\BBCQ}
\newblock
\APACjournalVolNumPages{Journal of Machine Learning
  Research}{12}{}{2825--2830}.
\PrintBackRefs{\CurrentBib}

\bibitem [\protect \citeauthoryear {%
{Pulkkinen}%
, {Bernabeu}%
, {Eichner}%
, {Viljanen}%
\BCBL {}\ \BBA {} {Ngwira}%
}{%
{Pulkkinen}%
\ \protect \BOthers {.}}{%
{\protect \APACyear {2015}}%
}]{%
2015_Pulkinnen_lengthscale}
\APACinsertmetastar {%
2015_Pulkinnen_lengthscale}%
\begin{APACrefauthors}%
{Pulkkinen}, A.%
, {Bernabeu}, E.%
, {Eichner}, J.%
, {Viljanen}, A.%
\BCBL {}\ \BBA {} {Ngwira}, C.%
\end{APACrefauthors}%
\unskip\
\newblock
\APACrefYearMonthDay{2015}{{\APACmonth{06}}}{}.
\newblock
{\BBOQ}\APACrefatitle {{Regional-scale high-latitude extreme geoelectric fields
  pertaining to geomagnetically induced currents}} {{Regional-scale
  high-latitude extreme geoelectric fields pertaining to geomagnetically
  induced currents}}.{\BBCQ}
\newblock
\APACjournalVolNumPages{Earth, Planets and Space}{67}{}{93}.
\newblock
\begin{APACrefDOI} \doi{10.1186/s40623-015-0255-6} \end{APACrefDOI}
\PrintBackRefs{\CurrentBib}

\bibitem [\protect \citeauthoryear {%
Pulkkinen%
, Rastatter%
\BCBL {}\ \protect \BOthers {.}}{%
Pulkkinen%
, Rastatter%
\BCBL {}\ \protect \BOthers {.}}{%
{\protect \APACyear {2013}}%
}]{%
pulkkinen2013community}
\APACinsertmetastar {%
pulkkinen2013community}%
\begin{APACrefauthors}%
Pulkkinen, A.%
, Rastatter, L.%
, Kuznetsova, M.%
, Singer, H.%
, Balch, C.%
, Weimer, D.%
\BDBL {}others%
\end{APACrefauthors}%
\unskip\
\newblock
\APACrefYearMonthDay{2013}{}{}.
\newblock
{\BBOQ}\APACrefatitle {Community-Wide Validation of Geospace Model Ground
  Magnetic Field Perturbation Predictions to Support Model Transition to
  Operations} {Community-wide validation of geospace model ground magnetic
  field perturbation predictions to support model transition to
  operations}.{\BBCQ}
\newblock
\APACjournalVolNumPages{Space Weather-the International Journal of Research and
  Applications}{11}{6}{369--385}.
\PrintBackRefs{\CurrentBib}

\bibitem [\protect \citeauthoryear {%
Pulkkinen%
, Rastätter%
\BCBL {}\ \protect \BOthers {.}}{%
Pulkkinen%
, Rastätter%
\BCBL {}\ \protect \BOthers {.}}{%
{\protect \APACyear {2013}}%
}]{%
Pulkkinen_metrics}
\APACinsertmetastar {%
Pulkkinen_metrics}%
\begin{APACrefauthors}%
Pulkkinen, A.%
, Rastätter, L.%
, Kuznetsova, M.%
, Singer, H.%
, Balch, C.%
, Weimer, D.%
\BDBL {}Weigel, R.%
\end{APACrefauthors}%
\unskip\
\newblock
\APACrefYearMonthDay{2013}{}{}.
\newblock
{\BBOQ}\APACrefatitle {Community-wide validation of geospace model ground
  magnetic field perturbation predictions to support model transition to
  operations} {Community-wide validation of geospace model ground magnetic
  field perturbation predictions to support model transition to
  operations}.{\BBCQ}
\newblock
\APACjournalVolNumPages{Space Weather}{11}{6}{369-385}.
\newblock
\begin{APACrefURL}
  \url{https://agupubs.onlinelibrary.wiley.com/doi/abs/10.1002/swe.20056}
  \end{APACrefURL}
\newblock
\begin{APACrefDOI} \doi{https://doi.org/10.1002/swe.20056} \end{APACrefDOI}
\PrintBackRefs{\CurrentBib}

\bibitem [\protect \citeauthoryear {%
Pulkkinen%
, Viljanen%
, Pajunpää%
\BCBL {}\ \BBA {} Pirjola%
}{%
Pulkkinen%
\ \protect \BOthers {.}}{%
{\protect \APACyear {2001}}%
}]{%
pulkkinen-2001}
\APACinsertmetastar {%
pulkkinen-2001}%
\begin{APACrefauthors}%
Pulkkinen, A.%
, Viljanen, A.%
, Pajunpää, K.%
\BCBL {}\ \BBA {} Pirjola, R.%
\end{APACrefauthors}%
\unskip\
\newblock
\APACrefYearMonthDay{2001}{}{}.
\newblock
{\BBOQ}\APACrefatitle {Recordings and occurrence of geomagnetically induced
  currents in the Finnish natural gas pipeline network} {Recordings and
  occurrence of geomagnetically induced currents in the finnish natural gas
  pipeline network}.{\BBCQ}
\newblock
\APACjournalVolNumPages{Journal of Applied Geophysics}{48}{4}{219-231}.
\newblock
\begin{APACrefURL}
  \url{https://www.sciencedirect.com/science/article/pii/S0926985101001082}
  \end{APACrefURL}
\newblock
\begin{APACrefDOI} \doi{https://doi.org/10.1016/S0926-9851(01)00108-2}
  \end{APACrefDOI}
\PrintBackRefs{\CurrentBib}

\bibitem [\protect \citeauthoryear {%
Rumelhart%
, Hinton%
\BCBL {}\ \BBA {} Williams%
}{%
Rumelhart%
\ \protect \BOthers {.}}{%
{\protect \APACyear {1985}}%
}]{%
rumelhart_1985_rnn}
\APACinsertmetastar {%
rumelhart_1985_rnn}%
\begin{APACrefauthors}%
Rumelhart, D\BPBI E.%
, Hinton, G\BPBI E.%
\BCBL {}\ \BBA {} Williams, R\BPBI J.%
\end{APACrefauthors}%
\unskip\
\newblock
\APACrefYearMonthDay{1985}{}{}.
\newblock
\APACrefbtitle {Learning internal representations by error propagation}
  {Learning internal representations by error propagation}\
  \APACbVolEdTR{}{\BTR{}}.
\newblock
\APACaddressInstitution{}{California Univ San Diego La Jolla Inst for Cognitive
  Science}.
\PrintBackRefs{\CurrentBib}

\bibitem [\protect \citeauthoryear {%
{Schrijver}%
, {Dobbins}%
, {Murtagh}%
\BCBL {}\ \BBA {} {Petrinec}%
}{%
{Schrijver}%
\ \protect \BOthers {.}}{%
{\protect \APACyear {2014}}%
}]{%
Schrijver:2014}
\APACinsertmetastar {%
Schrijver:2014}%
\begin{APACrefauthors}%
{Schrijver}, C\BPBI J.%
, {Dobbins}, R.%
, {Murtagh}, W.%
\BCBL {}\ \BBA {} {Petrinec}, S\BPBI M.%
\end{APACrefauthors}%
\unskip\
\newblock
\APACrefYearMonthDay{2014}{{\APACmonth{07}}}{}.
\newblock
{\BBOQ}\APACrefatitle {{Assessing the impact of space weather on the electric
  power grid based on insurance claims for industrial electrical equipment}}
  {{Assessing the impact of space weather on the electric power grid based on
  insurance claims for industrial electrical equipment}}.{\BBCQ}
\newblock
\APACjournalVolNumPages{Space Weather}{12}{7}{487-498}.
\newblock
\begin{APACrefDOI} \doi{10.1002/2014SW001066} \end{APACrefDOI}
\PrintBackRefs{\CurrentBib}

\bibitem [\protect \citeauthoryear {%
Smith%
\ \protect \BOthers {.}}{%
Smith%
\ \protect \BOthers {.}}{%
{\protect \APACyear {2020}}%
}]{%
2020_Smith_includeindices}
\APACinsertmetastar {%
2020_Smith_includeindices}%
\begin{APACrefauthors}%
Smith, A\BPBI W.%
, Rae, I\BPBI J.%
, Forsyth, C.%
, Oliveira, D\BPBI M.%
, Freeman, M\BPBI P.%
\BCBL {}\ \BBA {} Jackson, D\BPBI R.%
\end{APACrefauthors}%
\unskip\
\newblock
\APACrefYearMonthDay{2020}{}{}.
\newblock
{\BBOQ}\APACrefatitle {Probabilistic Forecasts of Storm Sudden Commencements
  From Interplanetary Shocks Using Machine Learning} {Probabilistic forecasts
  of storm sudden commencements from interplanetary shocks using machine
  learning}.{\BBCQ}
\newblock
\APACjournalVolNumPages{Space Weather}{18}{11}{e2020SW002603}.
\newblock
\begin{APACrefURL}
  \url{https://agupubs.onlinelibrary.wiley.com/doi/abs/10.1029/2020SW002603}
  \end{APACrefURL}
\newblock
\APACrefnote{e2020SW002603 10.1029/2020SW002603}
\newblock
\begin{APACrefDOI} \doi{https://doi.org/10.1029/2020SW002603} \end{APACrefDOI}
\PrintBackRefs{\CurrentBib}

\bibitem [\protect \citeauthoryear {%
Srivastava%
, Hinton%
, Krizhevsky%
, Sutskever%
\BCBL {}\ \BBA {} Salakhutdinov%
}{%
Srivastava%
\ \protect \BOthers {.}}{%
{\protect \APACyear {2014}}%
}]{%
srivastava2014dropout}
\APACinsertmetastar {%
srivastava2014dropout}%
\begin{APACrefauthors}%
Srivastava, N.%
, Hinton, G.%
, Krizhevsky, A.%
, Sutskever, I.%
\BCBL {}\ \BBA {} Salakhutdinov, R.%
\end{APACrefauthors}%
\unskip\
\newblock
\APACrefYearMonthDay{2014}{}{}.
\newblock
{\BBOQ}\APACrefatitle {Dropout: a simple way to prevent neural networks from
  overfitting} {Dropout: a simple way to prevent neural networks from
  overfitting}.{\BBCQ}
\newblock
\APACjournalVolNumPages{The journal of machine learning
  research}{15}{1}{1929--1958}.
\PrintBackRefs{\CurrentBib}

\bibitem [\protect \citeauthoryear {%
Tapping%
}{%
Tapping%
}{%
{\protect \APACyear {2013}}%
}]{%
F107_tapping}
\APACinsertmetastar {%
F107_tapping}%
\begin{APACrefauthors}%
Tapping, K\BPBI F.%
\end{APACrefauthors}%
\unskip\
\newblock
\APACrefYearMonthDay{2013}{}{}.
\newblock
{\BBOQ}\APACrefatitle {The 10.7 cm solar radio flux (F10.7)} {The 10.7 cm solar
  radio flux (f10.7)}.{\BBCQ}
\newblock
\APACjournalVolNumPages{Space Weather}{11}{7}{394-406}.
\newblock
\begin{APACrefURL}
  \url{https://agupubs.onlinelibrary.wiley.com/doi/abs/10.1002/swe.20064}
  \end{APACrefURL}
\newblock
\begin{APACrefDOI} \doi{https://doi.org/10.1002/swe.20064} \end{APACrefDOI}
\PrintBackRefs{\CurrentBib}

\bibitem [\protect \citeauthoryear {%
{T{\'o}th}%
\ \protect \BOthers {.}}{%
{T{\'o}th}%
\ \protect \BOthers {.}}{%
{\protect \APACyear {2005}}%
}]{%
toth_2005_swmf}
\APACinsertmetastar {%
toth_2005_swmf}%
\begin{APACrefauthors}%
{T{\'o}th}, G.%
, {Sokolov}, I\BPBI V.%
, {Gombosi}, T\BPBI I.%
, {Chesney}, D\BPBI R.%
, {Clauer}, C\BPBI R.%
, {de Zeeuw}, D\BPBI L.%
\BDBL {}{K{\'o}ta}, J.%
\end{APACrefauthors}%
\unskip\
\newblock
\APACrefYearMonthDay{2005}{{\APACmonth{12}}}{}.
\newblock
{\BBOQ}\APACrefatitle {{Space Weather Modeling Framework: A new tool for the
  space science community}} {{Space Weather Modeling Framework: A new tool for
  the space science community}}.{\BBCQ}
\newblock
\APACjournalVolNumPages{Journal of Geophysical Research (Space
  Physics)}{110}{A12}{A12226}.
\newblock
\begin{APACrefDOI} \doi{10.1029/2005JA011126} \end{APACrefDOI}
\PrintBackRefs{\CurrentBib}

\bibitem [\protect \citeauthoryear {%
{T{\'o}th}%
, {van der Holst}%
\BCBL {}\ \BBA {} {Huang}%
}{%
{T{\'o}th}%
\ \protect \BOthers {.}}{%
{\protect \APACyear {2011}}%
}]{%
toth_2011_swmf}
\APACinsertmetastar {%
toth_2011_swmf}%
\begin{APACrefauthors}%
{T{\'o}th}, G.%
, {van der Holst}, B.%
\BCBL {}\ \BBA {} {Huang}, Z.%
\end{APACrefauthors}%
\unskip\
\newblock
\APACrefYearMonthDay{2011}{{\APACmonth{05}}}{}.
\newblock
{\BBOQ}\APACrefatitle {{Obtaining Potential Field Solutions with Spherical
  Harmonics and Finite Differences}} {{Obtaining Potential Field Solutions with
  Spherical Harmonics and Finite Differences}}.{\BBCQ}
\newblock
\APACjournalVolNumPages{apj}{732}{2}{102}.
\newblock
\begin{APACrefDOI} \doi{10.1088/0004-637X/732/2/102} \end{APACrefDOI}
\PrintBackRefs{\CurrentBib}

\bibitem [\protect \citeauthoryear {%
{T{\'o}th}%
\ \protect \BOthers {.}}{%
{T{\'o}th}%
\ \protect \BOthers {.}}{%
{\protect \APACyear {2012}}%
}]{%
toth_2012_swmf}
\APACinsertmetastar {%
toth_2012_swmf}%
\begin{APACrefauthors}%
{T{\'o}th}, G.%
, {van der Holst}, B.%
, {Sokolov}, I\BPBI V.%
, {De Zeeuw}, D\BPBI L.%
, {Gombosi}, T\BPBI I.%
, {Fang}, F.%
\BDBL {}{Opher}, M.%
\end{APACrefauthors}%
\unskip\
\newblock
\APACrefYearMonthDay{2012}{{\APACmonth{02}}}{}.
\newblock
{\BBOQ}\APACrefatitle {{Adaptive numerical algorithms in space weather
  modeling}} {{Adaptive numerical algorithms in space weather
  modeling}}.{\BBCQ}
\newblock
\APACjournalVolNumPages{Journal of Computational Physics}{231}{3}{870-903}.
\newblock
\begin{APACrefDOI} \doi{10.1016/j.jcp.2011.02.006} \end{APACrefDOI}
\PrintBackRefs{\CurrentBib}

\bibitem [\protect \citeauthoryear {%
UN%
}{%
UN%
}{%
{\protect \APACyear {2017}}%
}]{%
un_space}
\APACinsertmetastar {%
un_space}%
\begin{APACrefauthors}%
UN.%
\end{APACrefauthors}%
\unskip\
\newblock
\APACrefYearMonthDay{2017}{}{}.
\newblock
\APACrefbtitle {{United Nations Office of outer space affairs, International
  Spaceweather Initiative}.} {{United Nations Office of outer space affairs,
  International Spaceweather Initiative}.}
\newblock
\APAChowpublished
  {\url{https://www.unoosa.org/oosa/en/ourwork/psa/bssi/iswi.html}}.
\PrintBackRefs{\CurrentBib}

\bibitem [\protect \citeauthoryear {%
{Upendran}%
, {Cheung}%
, {Hanasoge}%
\BCBL {}\ \BBA {} {Krishnamurthi}%
}{%
{Upendran}%
\ \protect \BOthers {.}}{%
{\protect \APACyear {2020}}%
}]{%
upendran_2020_solarwind}
\APACinsertmetastar {%
upendran_2020_solarwind}%
\begin{APACrefauthors}%
{Upendran}, V.%
, {Cheung}, M\BPBI C\BPBI M.%
, {Hanasoge}, S.%
\BCBL {}\ \BBA {} {Krishnamurthi}, G.%
\end{APACrefauthors}%
\unskip\
\newblock
\APACrefYearMonthDay{2020}{{\APACmonth{09}}}{}.
\newblock
{\BBOQ}\APACrefatitle {{Solar Wind Prediction Using Deep Learning}} {{Solar
  Wind Prediction Using Deep Learning}}.{\BBCQ}
\newblock
\APACjournalVolNumPages{Space Weather}{18}{9}{e02478}.
\newblock
\begin{APACrefDOI} \doi{10.1029/2020SW002478} \end{APACrefDOI}
\PrintBackRefs{\CurrentBib}

\bibitem [\protect \citeauthoryear {%
Upendran%
\ \protect \BOthers {.}}{%
Upendran%
\ \protect \BOthers {.}}{%
{\protect \APACyear {2022}}%
}]{%
Daggermodel_github}
\APACinsertmetastar {%
Daggermodel_github}%
\begin{APACrefauthors}%
Upendran, V.%
, Tigas, P.%
, Ferdousi, B.%
, Bloch, T.%
, Cheung, M\BPBI C\BPBI M.%
, Ganju, S.%
\BDBL {}Gal, Y.%
\end{APACrefauthors}%
\unskip\
\newblock
\APACrefYearMonthDay{2022}{{\APACmonth{04}}}{}.
\newblock
\APACrefbtitle {Vishal-Upendran/geoeffectivenet-1: DAGGER model.}
  {Vishal-upendran/geoeffectivenet-1: Dagger model.}
\newblock
\APACaddressPublisher{}{Zenodo}.
\newblock
\begin{APACrefURL} \url{https://doi.org/10.5281/zenodo.6410499}
  \end{APACrefURL}
\newblock
\begin{APACrefDOI} \doi{10.5281/zenodo.6410499} \end{APACrefDOI}
\PrintBackRefs{\CurrentBib}

\bibitem [\protect \citeauthoryear {%
{Verbanac}%
, {Mandea}%
, {Vr{\v{s}}nak}%
\BCBL {}\ \BBA {} {Sentic}%
}{%
{Verbanac}%
\ \protect \BOthers {.}}{%
{\protect \APACyear {2011}}%
}]{%
Verbanac_solarGeomagCorr}
\APACinsertmetastar {%
Verbanac_solarGeomagCorr}%
\begin{APACrefauthors}%
{Verbanac}, G.%
, {Mandea}, M.%
, {Vr{\v{s}}nak}, B.%
\BCBL {}\ \BBA {} {Sentic}, S.%
\end{APACrefauthors}%
\unskip\
\newblock
\APACrefYearMonthDay{2011}{{\APACmonth{07}}}{}.
\newblock
{\BBOQ}\APACrefatitle {{Evolution of Solar and Geomagnetic Activity Indices,
  and Their Relationship: 1960 - 2001}} {{Evolution of Solar and Geomagnetic
  Activity Indices, and Their Relationship: 1960 - 2001}}.{\BBCQ}
\newblock
\APACjournalVolNumPages{Solar physics}{271}{1-2}{183-195}.
\newblock
\begin{APACrefDOI} \doi{10.1007/s11207-011-9801-y} \end{APACrefDOI}
\PrintBackRefs{\CurrentBib}

\bibitem [\protect \citeauthoryear {%
Virtanen%
\ \protect \BOthers {.}}{%
Virtanen%
\ \protect \BOthers {.}}{%
{\protect \APACyear {2020}}%
}]{%
scipy}
\APACinsertmetastar {%
scipy}%
\begin{APACrefauthors}%
Virtanen, P.%
, Gommers, R.%
, Oliphant, T\BPBI E.%
, Haberland, M.%
, Reddy, T.%
, Cournapeau, D.%
\BDBL {}{SciPy 1.0 Contributors}%
\end{APACrefauthors}%
\unskip\
\newblock
\APACrefYearMonthDay{2020}{}{}.
\newblock
{\BBOQ}\APACrefatitle {{{SciPy} 1.0: Fundamental Algorithms for Scientific
  Computing in Python}} {{{SciPy} 1.0: Fundamental Algorithms for Scientific
  Computing in Python}}.{\BBCQ}
\newblock
\APACjournalVolNumPages{Nature Methods}{17}{}{261--272}.
\newblock
\begin{APACrefDOI} \doi{10.1038/s41592-019-0686-2} \end{APACrefDOI}
\PrintBackRefs{\CurrentBib}

\bibitem [\protect \citeauthoryear {%
Weigel%
, Vassiliadis%
\BCBL {}\ \BBA {} Klimas%
}{%
Weigel%
\ \protect \BOthers {.}}{%
{\protect \APACyear {2002}}%
}]{%
weigel-2002}
\APACinsertmetastar {%
weigel-2002}%
\begin{APACrefauthors}%
Weigel, R\BPBI S.%
, Vassiliadis, D.%
\BCBL {}\ \BBA {} Klimas, A\BPBI J.%
\end{APACrefauthors}%
\unskip\
\newblock
\APACrefYearMonthDay{2002}{}{}.
\newblock
{\BBOQ}\APACrefatitle {Coupling of the solar wind to temporal fluctuations in
  ground magnetic fields} {Coupling of the solar wind to temporal fluctuations
  in ground magnetic fields}.{\BBCQ}
\newblock
\APACjournalVolNumPages{Geophysical Research Letters}{29}{19}{21-1-21-4}.
\newblock
\begin{APACrefURL}
  \url{https://agupubs.onlinelibrary.wiley.com/doi/abs/10.1029/2002GL014740}
  \end{APACrefURL}
\newblock
\begin{APACrefDOI} \doi{https://doi.org/10.1029/2002GL014740} \end{APACrefDOI}
\PrintBackRefs{\CurrentBib}

\bibitem [\protect \citeauthoryear {%
D.~Weimer%
\ \protect \BOthers {.}}{%
D.~Weimer%
\ \protect \BOthers {.}}{%
{\protect \APACyear {2010}}%
}]{%
weimer2010statistical}
\APACinsertmetastar {%
weimer2010statistical}%
\begin{APACrefauthors}%
Weimer, D.%
, Clauer, C.%
, Engebretson, M.%
, Hansen, T.%
, Gleisner, H.%
, Mann, I.%
\BCBL {}\ \BBA {} Yumoto, K.%
\end{APACrefauthors}%
\unskip\
\newblock
\APACrefYearMonthDay{2010}{}{}.
\newblock
{\BBOQ}\APACrefatitle {Statistical Maps of Geomagnetic Perturbations as a
  Function of the Interplanetary Magnetic Field} {Statistical maps of
  geomagnetic perturbations as a function of the interplanetary magnetic
  field}.{\BBCQ}
\newblock
\APACjournalVolNumPages{Journal of Geophysical Research: Space
  Physics}{115}{A10}{}.
\PrintBackRefs{\CurrentBib}

\bibitem [\protect \citeauthoryear {%
D\BPBI R.~Weimer%
}{%
D\BPBI R.~Weimer%
}{%
{\protect \APACyear {2013}}%
}]{%
weimer2013empirical}
\APACinsertmetastar {%
weimer2013empirical}%
\begin{APACrefauthors}%
Weimer, D\BPBI R.%
\end{APACrefauthors}%
\unskip\
\newblock
\APACrefYearMonthDay{2013}{}{}.
\newblock
{\BBOQ}\APACrefatitle {An Empirical Model of Ground-Level Geomagnetic
  Perturbations} {An empirical model of ground-level geomagnetic
  perturbations}.{\BBCQ}
\newblock
\APACjournalVolNumPages{Space Weather-the International Journal of Research and
  Applications}{11}{3}{107--120}.
\PrintBackRefs{\CurrentBib}

\bibitem [\protect \citeauthoryear {%
{Welling}%
, {Anderson}%
, {Crowley}%
, {Pulkkinen}%
\BCBL {}\ \BBA {} {Rast{\"a}tter}%
}{%
{Welling}%
\ \protect \BOthers {.}}{%
{\protect \APACyear {2017}}%
}]{%
welling-evaluations-2017}
\APACinsertmetastar {%
welling-evaluations-2017}%
\begin{APACrefauthors}%
{Welling}, D\BPBI T.%
, {Anderson}, B\BPBI J.%
, {Crowley}, G.%
, {Pulkkinen}, A\BPBI A.%
\BCBL {}\ \BBA {} {Rast{\"a}tter}, L.%
\end{APACrefauthors}%
\unskip\
\newblock
\APACrefYearMonthDay{2017}{{\APACmonth{01}}}{}.
\newblock
{\BBOQ}\APACrefatitle {{Exploring predictive performance: A reanalysis of the
  geospace model transition challenge}} {{Exploring predictive performance: A
  reanalysis of the geospace model transition challenge}}.{\BBCQ}
\newblock
\APACjournalVolNumPages{Space Weather}{15}{1}{192-203}.
\newblock
\begin{APACrefDOI} \doi{10.1002/2016SW001505} \end{APACrefDOI}
\PrintBackRefs{\CurrentBib}

\bibitem [\protect \citeauthoryear {%
Welling%
\ \protect \BOthers {.}}{%
Welling%
\ \protect \BOthers {.}}{%
{\protect \APACyear {2018}}%
}]{%
welling_2018_metrics}
\APACinsertmetastar {%
welling_2018_metrics}%
\begin{APACrefauthors}%
Welling, D\BPBI T.%
, Ngwira, C\BPBI M.%
, Opgenoorth, H.%
, Haiducek, J\BPBI D.%
, Savani, N\BPBI P.%
, Morley, S\BPBI K.%
\BDBL {}Liemohn, M.%
\end{APACrefauthors}%
\unskip\
\newblock
\APACrefYearMonthDay{2018}{}{}.
\newblock
{\BBOQ}\APACrefatitle {Recommendations for Next-Generation Ground Magnetic
  Perturbation Validation} {Recommendations for next-generation ground magnetic
  perturbation validation}.{\BBCQ}
\newblock
\APACjournalVolNumPages{Space Weather}{16}{12}{1912-1920}.
\newblock
\begin{APACrefURL}
  \url{https://agupubs.onlinelibrary.wiley.com/doi/abs/10.1029/2018SW002064}
  \end{APACrefURL}
\newblock
\begin{APACrefDOI} \doi{https://doi.org/10.1029/2018SW002064} \end{APACrefDOI}
\PrintBackRefs{\CurrentBib}

\bibitem [\protect \citeauthoryear {%
{Wintoft}%
, {Wik}%
\BCBL {}\ \BBA {} {Viljanen}%
}{%
{Wintoft}%
\ \protect \BOthers {.}}{%
{\protect \APACyear {2015}}%
}]{%
wintoft-dbdt-model}
\APACinsertmetastar {%
wintoft-dbdt-model}%
\begin{APACrefauthors}%
{Wintoft}, P.%
, {Wik}, M.%
\BCBL {}\ \BBA {} {Viljanen}, A.%
\end{APACrefauthors}%
\unskip\
\newblock
\APACrefYearMonthDay{2015}{{\APACmonth{03}}}{}.
\newblock
{\BBOQ}\APACrefatitle {{Solar wind driven empirical forecast models of the time
  derivative of the ground magnetic field}} {{Solar wind driven empirical
  forecast models of the time derivative of the ground magnetic field}}.{\BBCQ}
\newblock
\APACjournalVolNumPages{Journal of Space Weather and Space Climate}{5}{}{A7}.
\newblock
\begin{APACrefDOI} \doi{10.1051/swsc/2015008} \end{APACrefDOI}
\PrintBackRefs{\CurrentBib}

\end{thebibliography}

%Reference citation instructions and examples:
%
% Please use ONLY \cite and \citeA for reference citations.
% \cite for parenthetical references
% ...as shown in recent studies (Simpson et al., 2019)
% \citeA for in-text citations
% ...Simpson et al. (2019) have shown...
%
%
%...as shown by \citeA{jskilby}.
%...as shown by \citeA{lewin76}, \citeA{carson86}, \citeA{bartoldy02}, and \citeA{rinaldi03}.
%...has been shown \cite{jskilbye}.
%...has been shown \cite{lewin76,carson86,bartoldy02,rinaldi03}.
%... \cite <i.e.>[]{lewin76,carson86,bartoldy02,rinaldi03}.
%...has been shown by \cite <e.g.,>[and others]{lewin76}.
%
% apacite uses < > for prenotes and [ ] for postnotes
% DO NOT use other cite commands (e.g., \citet, \citep, \citeyear, \nocite, \citealp, etc.).
%

\end{document}